\documentclass[
  12pt,
  nofootinbib,
  longbibliography,
  aps,
  prX,
  superscriptaddress,
  floatfix,
  showpacs,
  notitlepage,
]{revtex4-1}
\usepackage{amsmath,amsthm,amssymb,color,psfrag,url,latexsym,graphicx,epstopdf,slashed,xspace,hyperref,enumitem}

\usepackage{lineno}

\definecolor{darkred}{rgb}{0.6,0.0,0.0}
\definecolor{darkblue}{rgb}{0.0,0.0,0.5}
\definecolor{darkgreen}{rgb}{0.0,0.5,0.0}
\definecolor{brown}{rgb}{0.0,0.0,0.0}

\newcommand{\be}{\begin{equation}}
\newcommand{\ee}{\end{equation}}
\newcommand{\bea}{\begin{eqnarray}}
\newcommand{\eea}{\end{eqnarray}}

\newcommand{\alphas}{\alpha_{S}}
\newcommand{\alphasmZ}{\alphas(m^{2}_\mathrm{Z})}
\newcommand{\epem}{{\rm e^+e^-}}
\newcommand{\MSbar}{\overline{\mathrm{MS}}}

\begin{document}
\title{Opportunities for precision QCD physics in hadronization at Belle~II -- a Snowmass whitepaper}

\newcommand*{\LBL}{Lawrence Berkeley National Laboratory, Berkeley, CA 94720, USA} 
\newcommand*{\Durham}{Institute for Particle Physics Phenomenology, Durham University, United Kingdom}
\newcommand*{\SBU}{Stony Brook, State University of New York, NY 11794, USA}
\newcommand*{\JLAB}{Thomas Jefferson National Accelerator Facility, Newport News, VA 23606, USA}
\newcommand*{\TEMPLE}{Temple University, Philadelphia, PA 19122, USA}
\newcommand*{\ANL}{Argonne National Laboratory, Lemont, IL 60439, USA}
\newcommand*{\MIT}{Massachusetts Institute of Technology, Cambridge, MA 02139, USA}
\newcommand*{\WM}{The College of William and Mary, Williamsburg, VA 23185, USA}
\newcommand*{\DUKE}{Duke University, Durham, NC 27708, USA} 
\newcommand*{\Hampton}{Hampton University, Hampton, VA 23668, USA}
\newcommand*{\UCLA}{Department of Physics and Astronomy, University of California, Los Angeles, CA, USA}
\newcommand*{\UCLATheory}{Mani L. Bhaumik Institute for Theoretical Physics, University of California, Los Angeles, CA, USA}
\newcommand*{\LANL}{Los Alamos National Laboratory, ....} 
\newcommand*{\CNF}{Center for Nuclear Femtography, 1201 New York Avenue, Washington, DC 20005, USA}
\newcommand*{\CFNS}{Center for Frontiers in Nuclear Science, Stony Brook University, Stony Brook, NY, USA}
\newcommand*{\CEA}{IRFU CEA-Saclay, 91191 Gif sur Yvette, France}
\newcommand*{\ORSAY}{Université Paris-Sud 11, Orsay, France} 
\newcommand*{\FER}{INFN Ferrara, 44122 Ferrara, Italy}
\newcommand*{\KNU}{Department of Physics, Kyungpook National University, Daegu 41566, Korea}
\newcommand*{\UMD}{University of Maryland, College Park, MD 20742, USA}
\newcommand*{\BNL}{Department of Physics, Brookhaven National Laboratory, Upton, NY 11973, USA}
\newcommand*{\Regensburg}{Institut f\"ur Theoretische Physik, Universit\"at Regensburg, D-93040 Regensburg, Germany}
\newcommand*{\Regina}{University of Regina, Regina, SK S4S~0A2 Canada}
\newcommand*{\UVA}{University of Virginia, Physics Department, 382 McCormick Rd, Charlottesville, VA 22904, USA}
\newcommand*{\FIU}{Florida International University, Miami, FL, USA}
\newcommand*{\Glasgow}{SUPA, School of Physics and Astronomy, University of Glasgow, Glasgow, G12 8QQ, United Kingdom}
\newcommand*{\IUB}{Indiana University, Bloomington, IN 47408, USA}
\newcommand*{\ODU}{Old Dominion University, Department of Physics, 4600 Elkhorn Ave., Norfolk, VA 23529, USA}
\newcommand*{\GWU}{The George Washington University, Washington, DC 20052, USA} 
\newcommand*{\CAM}{DAMTP, University of Cambridge, United Kingdom}
\newcommand*{\APCTP}{Asia Pacific Center for Theoretical Physics, Pohang, Gyeongbuk 37673, Korea}
\newcommand*{\Zagreb}{University of Zagreb, Devision of Theoretical Physics, Croatia}
\newcommand*{\Catania}{INFN Sezione di Catania, I-95123 Catania, Italy}
\newcommand*{\Messina}{Dipartimento di Scienze Matematiche e Informatiche, Scienze Fisiche e Scienze della Terra,\\ Universita` degli Studi di Messina, I-98122 Messina, Italy}
\newcommand*{\CNRS}{Laboratoire de Physique Joliot-Curie, CNRS-IN2P3, Universit\'e Paris-Saclay, France}
\newcommand*{\RIKEN}{RIKEN Nishina Center for Accelerator-Based Science, Wako, Saitama 351-0198, Japan}
\newcommand*{\Bilbao}{Department of Physics \& EHU Quantum Center, University of the Basque Country UPV/EHU, 48080 Bilbao, Spain}
\newcommand*{\IkerB}{IKERBASQUE, Basque Foundation for Science, 48013 Bilbao, Spain}
\newcommand*{\RikenBNL}{RIKEN BNL Research Center, Brookhaven National Laboratory, Upton, NY 11973-5000, USA}
\newcommand*{\UNAM}{Instituto de F\'{i}sica, Universidad Nacional Aut\'{o}noma de M\'{e}xico, Apartado Postal 20-364,\\ 01000 Ciudad de M\'{e}xico, Mexico}
\newcommand*{\BERKS}{Science Division, Penn State University Berks, Reading, PA 19610, USA}
\newcommand*{\UPAVIA}{Dipartimento di Fisica, Universit\`{a} di Pavia, Pavia, Italy}
\newcommand*{\INFNPAVIA}{INFN, Sezione di Pavia, Pavia, Italy}
\newcommand*{\CNYANG}{C.N. Yang Institute for Theoretical Physics, Stony Brook University, Stony Brook, NY 11794, USA}
\newcommand*{\GSU}{Physics and Astronomy Department, Georgia State University, Atlanta, GA 30303, USA}
\newcommand*{\MainzU}{Institut für Physik, Institut für Kernphysik, Johannes-Gutenberg-Universität, D-55099 Mainz, Germany}
\newcommand*{\Madrid}{Departamento de F\'{i}sica Te\'{o}rica \& IPARCOS, Facultad de Ciencias F\'{i}sicas, Universidad Complutense de Madrid, E-28040 Madrid, Spain}
\newcommand*{\Vienna}{Wien Universit\"at, Faculty of Physics, Boltzmanngasse 5, A-1090 Vienna, Austria}
\newcommand*{\Lund}{Department of Astronomy and Theoretical Physics, Lund University,
Lund, Sweden}
\newcommand*{\Barcelona}{Institut de F{\'i}sica d’Altes Energies (IFAE) and Barcelona Institute of Science and Technology (BIST), Campus UAB, 08193 Bellaterra, Spain} 
\newcommand*{\CERN}{CERN, EP Department, CH-1211 Geneva 23, Switzerland}

\author{A.~Accardi}
\affiliation{\Hampton}
\affiliation{\JLAB}
\author{Y.-T.~Chien}
\affiliation{\SBU}
\affiliation{\CNYANG}
\affiliation{\CFNS}
\affiliation{\GSU}

\author{D.~d'Enterria}
\affiliation{\CERN}
\author{A.~Deshpande}
\affiliation{\CFNS}
\affiliation{\SBU}
\affiliation{\BNL}
\author{C.~Dilks}
\affiliation{\DUKE}
\author{P.~A.~Gutierrez Garcia}
\affiliation{\Madrid}
\author{W.~W.~Jacobs}
\affiliation{\IUB}
\author{F.~Krauss}
\affiliation{\Durham}
\author{S.~Leal Gomez}
\affiliation{\Vienna}
\author{M.~Mouli Mondal}
\affiliation{\CFNS}
\affiliation{\SBU}
\author{K.~Parham}
\affiliation{\DUKE}
\author{F.~Ringer}
\affiliation{\CNYANG}
\author{P.~Sanchez-Puertas}
\affiliation{\Barcelona}
\author{S.~Schneider}
\affiliation{\DUKE}
\author{G.~Schnell}
\affiliation{\Bilbao}
\affiliation{\IkerB}
\author{I.~Scimemi}
\affiliation{\Madrid}
\author{R.~Seidl}
\affiliation{\RIKEN}
\affiliation{\RikenBNL}
\author{A.~Signori}
\affiliation{\UPAVIA}
\affiliation{\INFNPAVIA}
\author{T.~Sj\"ostrand}
\affiliation{\Lund}
\author{G.~Sterman}
\affiliation{\CFNS}
\affiliation{\SBU}
\affiliation{\CNYANG}
\author{A.~Vossen}
\thanks{editor, contact: \href{mailto:anselm.vossen@duke.edu}{anselm.vossen@duke.edu}
}
\affiliation{\DUKE}
\affiliation{\JLAB}

\begin{abstract}
    This document presents a selection of QCD studies accessible to high-precision studies with hadronic final states in $\epem$ collisions at Belle~II. The exceptionally clean environment and the state-of-the-art capabilities of the Belle~II detector (including excellent particle identification and improved vertex reconstruction), coupled with an unprecedented data-set size, will make possible to carry out multiple valuable measurements of the strong interaction including hadronic contributions to the muon $(g-2)$ and the QCD coupling, as well as advanced studies of parton hadronization and dynamical quark mass generation.
\end{abstract}

\maketitle
\tableofcontents
\section{Introduction}
The QCD part of the Lagrangian of the Standard Model describing strong interactions has, besides the quark masses, only one parameter, the coupling strength $\alpha_S$. Nevertheless, the dynamics of QCD is responsible for the complex properties of most of the visible mass in the universe, such as the mass and spin of nucleons. This has made investigations of QCD a central part of nuclear and particle physics over the last 50 years. 
This document lays out a program to study key QCD processes in $\epem$ hadronic final states with the Belle~II experiment at SuperKEKB. As will be evident in the following material, the exceptionally clean environment, the capabilities of the Belle~II detector, including excellent particle identification and improved vertex reconstruction capabilities, coupled with an unprecedented data-set size accumulated with up to 50~ab$^{-1}$ integrated luminosities~\cite{Belle-II:2018jsg}, will make unique studies of the strong interaction possible. 
The document is organized as follows. 
Section~\ref{sec:g-2} discusses the planned program at Belle~II to reduce the uncertainties from hadronic corrections to $g-2$.
Section~\ref{sec:alphas} summarizes the interesting opportunities for high-precision determinations of the strong coupling $\alphas$ from various hadronic final states in $\epem$ collisions at Belle~II.
Sec.~\ref{sec:hadronization} covers the program of QCD studies in hadronization in detail. Aspects investigated include fragmentation studies in Sec.~\ref{sec:hadronAndJet}, in particular with applications for physics programs at Jefferson Lab and the future Electron-Ion Collider, interrelation with Monte Carlo models in Sec.~\ref{sec:tuneMC}, with a special consideration for emerging generators taking polarization into account (Sec.~\ref{sec:polmc}), jet and transverse momentum dependent (TMD) physics in Sec.~\ref{sec:jets}, correlation studies to map out hadronization dynamics in Sec.~\ref{sec:correlations} and studies of dynamical mass generation in Sec~\ref{sec:jet_mass}. 


\section{Constraining systematic uncertainties on the determination of \lowercase{g -- 2}}
\label{sec:g-2}
One of the most promising smoking guns for physics beyond the Standard Model (SM) is the muon's anomalous magnetic moment $a_\mu$, the deviation of the gyromagnetic ratio $g$ from the Dirac value of 2, which is parameterized as $a_\mu= \frac{g_\mu-2}{2}$. Both the SM theoretical prediction~\cite{aoyama:2012wk,Aoyama:2019ryr,czarnecki:2002nt,gnendiger:2013pva,davier:2017zfy,keshavarzi:2018mgv,colangelo:2018mtw,hoferichter:2019mqg,davier:2019can,keshavarzi:2019abf,kurz:2014wya,melnikov:2003xd,masjuan:2017tvw,Colangelo:2017fiz,hoferichter:2018kwz,gerardin:2019vio,bijnens:2019ghy,colangelo:2019uex,Blum:2019ugy,colangelo:2014qya} as well as the experimental determination have reached astounding precision (cf. Ref.~\cite{Aoyama:2020ynm}). The current experimental value (combining the BNL E821 result~\cite{Muong-2:2006rrc} with the first result from the Fermilab g-2 experiment~\cite{Muong-2:2015xgu}) differs from the SM prediction by 4.2 sigma: $a_\mu^\text{exp} - a_\mu^\text{SM} = (251 \pm 59) \times 10^{-11}$~\cite{Muong-2:2021ojo}. While the QED/electroweak~\cite{czarnecki:2002nt,gnendiger:2013pva,aoyama:2012wk,Aoyama:2019ryr} contributions to $a_\mu $ are well under control in comparison to experimental uncertainties, QCD corrections dominate the theory uncertainty and are of similar size as current experimental uncertainties. The QCD corrections comprise the hadronic light-by-light scattering (HLbL) and the hadronic vacuum polarization (HVP) contributions, which are both still relatively poorly constrained, with HVP giving the larger contribution to the overall uncertainty. With the upcoming improvements from the experimental side, the theoretical uncertainties from hadronic contributions will become the main limiting factor. As such, a large community effort has formed to tackle those~\cite{Aoyama:2020ynm}. In the remainder, the focus will be on the largest contribution to the SM uncertainty, the HVP contributions. The importance of getting these contributions under control was demonstrated by new lattice calculations~\cite{Borsanyi:2020mff} for the HVP, which reduced the data--theory tension to about 2~$\sigma$ but which are in apparent conflict with experimental measurements.
While lattice calculations have developed into rather valuable and continuously improving tools, the gold standard for the determination of the HVP contribution is to use experimental data, in particular, hadron electroproduction in connection with dispersion relations, where Belle~II can play an important role. An alternative experimental approach, using the spectral function for $\tau\rightarrow \pi \pi^0 \nu$ and making a correction for possible isospin symmetry violations, is another process where Belle~II can contribute. 
Below, we discuss both approaches and detail how the Belle~II physics program will be vital for the determination of the HVP contributions but also for improving the understanding of HLbL contributions, and thus for the discovery potential of the g-2 and upcoming experiments.\footnote{A novel approach to tackle HVP in the space-like regime via $\mu$--$e$ elastic scattering has been proposed and developed as MUonE within the {\em Physics Beyond Collider} initiative at CERN~\cite{QCDWorkingGroup:2019dyv}.}

\subsection{Constraining HVP through $e^+e^-\rightarrow$ hadrons and dispersion relations}

The HVP contribution to $a_\mu$ can be constrained using the dispersion relation 
$a_\mu^{\mathrm{HVP,LO}}=\frac{\alpha^2}{3\pi^2}\int_{M_\pi^2}^\infty
\frac{K(s)}{s} R(s) ds$, where $K(s)$ is a slowly varying kernel function, $s$ the Mandelstam variable, and $R(s)$ is the so-called hadronic $R$ ratio, $R(s)=\frac{3s}{4\pi\alpha^2}\, \sigma_h(e^+e^-\rightarrow \textrm{hadrons})$~\cite{Aoyama:2020ynm}.
For the dispersion integral, $\sigma_h$ has to be evaluated over the full $s$ range. This can be accomplished by explicit beam-energy scans, or---as pursued at, e.g., the B-factories---using the technique of radiative return, thanks to the high luminosities of the B-factories~\cite{BaBar:2014omp}. The technique of radiative return takes advantage of initial-state radiation (ISR) to scan the partonic $s$. An ISR photon is detected, such that the effective $\sqrt{s}$ value is lowered by the energy of the photon. One disadvantage of this method is the accurate description and treatment of ISR and final-state radiation, including their interference, in the analysis by the experiments and phenomenology. Different experimental collaborations employ different and sometimes even multiple approaches, which needs to be factored in the posterior analysis and might give rise to tensions as perhaps the presently existing one between the BaBar~\cite{BaBar:2009wpw} and KLOE~\cite{KLOE-2:2017fda} data discussed below.

While the full $s$ range is required for an exact evaluation of $a_\mu^{\mathrm{HVP,LO}}$, the low-$s$ region is the most important contribution due to the $1/s$ factor in the integral.
Moreover, about $70$\% of the contribution to the dispersion integral comes from the channel $e^+e^-\rightarrow \pi\pi$, in particular, from the region around the $\rho$ and $\omega$ resonances.

\begin{figure}
    \centering
    \includegraphics[width=0.525\textwidth]{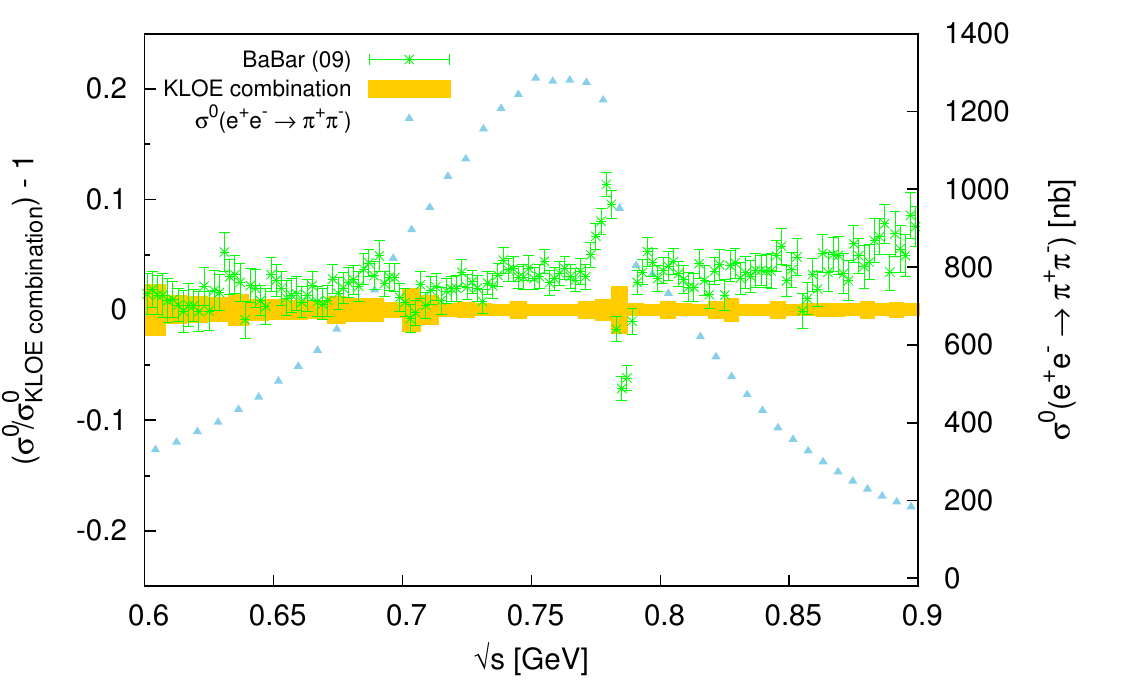}
    \includegraphics[width=0.465\textwidth]{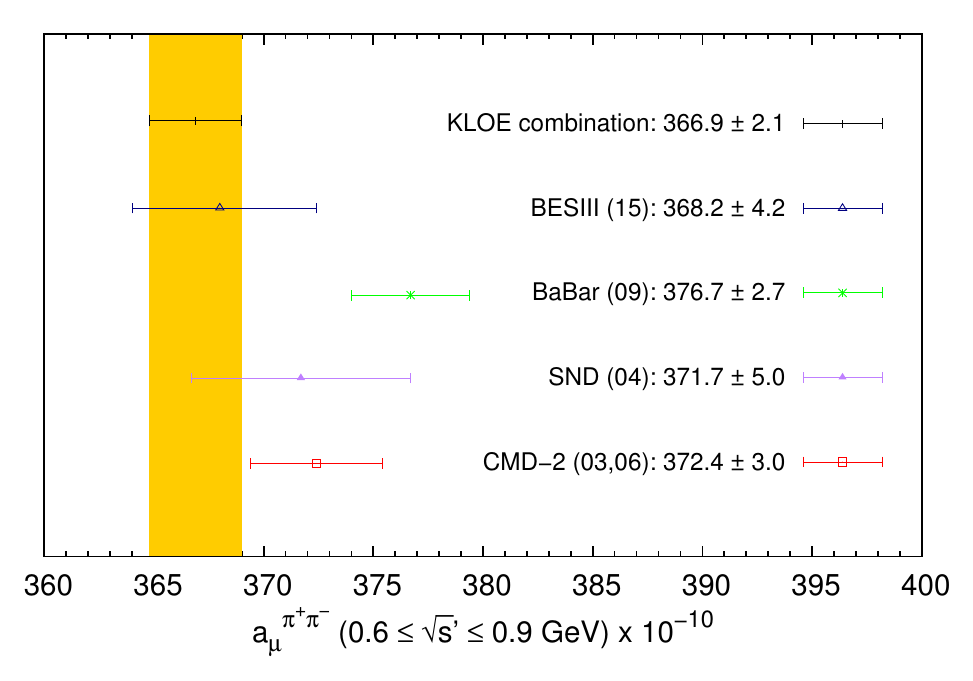}
    \caption{Illustrations of the current tension between BaBar~\cite{BaBar:2009wpw} and KLOE~\cite{KLOE-2:2017fda} data. (left) Ratio of the BaBar to KLOE data for the exclusive annihilation cross section into $\pi^+\pi^-$ pairs. (right) Comparison of the leading-order HVP contribution to $a_\mu$ from the (dominant) $\pi^+\pi^-$ channel using different (sub)sets of data. (Figures from Ref.~\cite{KLOE-2:2017fda})}
    \label{fig:BabarKloeHVPTension}
\end{figure}

Given the tremendous interest in $a_\mu$, a large effort has gone into the determination of the 2$\pi$ contribution to $R(s)$. Indeed, in many cases the measurements become systematics dominated. However, one of the main problems with the current data sets is the tension between the high-precision measurements from BaBar~\cite{BaBar:2009wpw} and KLOE~\cite{KLOE-2:2017fda}. Figure \ref{fig:BabarKloeHVPTension} illustrates the problems: the BaBar data in the left plot do not only overshoot the KLOE combination of the $e^+e^-\rightarrow \pi^+\pi^-$ cross section in the region of interest, the difference between the two experiments also increases with invariant mass of the pion pair. Striking is also the clear inconsistency in the $\rho$--$\omega$ interference region, though less problematic when performing the integral over $s$. The right plot of Fig.~\ref{fig:BabarKloeHVPTension} shows the contribution to $a_\mu^{\mathrm{HVP,LO}}$ from the two-pion channel, elucidating again the strong tension between the KLOE and BaBar data. Playing such an important role in the total HVP contribution to $a_\mu$, it is thus an issue of great urgency to not only reduce the uncertainty on the hadronic cross-section measurements as such to match the expected experimental uncertainty on $a_\mu$, but also to resolve this significant tension  between the measurements with the currently highest precision, KLOE and BaBar, as it currently constitutes a major contribution to the uncertainty on the SM prediction for $a_\mu$.

The tension translates in a difference of $a_\mu^\textrm{HVP}$ at LO of about $6\times 10^{-10}$~\cite{davier:2019can,Aoyama:2020ynm}. In addition to fundamental questions about the experimental data, this difference is of the same order of magnitude as the BNL E821 experimental precision. And its contribution to the SM estimate of $a_\mu$ will dominate over the aimed-for precision of the final $g-2$ result.
As should be clear from the above description, an effort to precisely measure $\sigma_h$, in particular in the $\rho$--$\omega$ interference region, is of the highest priority to the $g-2$ theory community to reduce the overall uncertainties to the level of the expected uncertainties of the final $g-2$ result and to resolve tension in the existing data. The supporting letter by the $g-2$ theory initiative attests to this fact. 

At Belle~II, the cross section for the $e^+e^-\rightarrow \pi^+\pi^-(\gamma)$ process can be measured from threshold to 3~GeV reduced centre-of-mass energy $(\sqrt{s'})$. Belle~II implemented a pure photon trigger enabling a close to 100\% efficiency for ISR events of interest, in contrast to what was in place at Belle. Even though the cross section for ISR emission at these lower $\sqrt{s}$ drops, the SuperKEKB luminosities will enable a high-precision measurement that from the current perspective will be dominated by systematic uncertainties. With the complete Belle~II data set, the expected statistical uncertainties are of the level of 5~ppm, or three times lower than expected experimental uncertainty for the final Fermilab g-2 result on $a_\mu$. 
The leading contributions to the systematic uncertainties are the efficiencies of $\pi$ detection as well as the $\pi/\mu$ particle identification (PID). Previous BaBar measurements reached a systematic uncertainty of 31~ppm, 1/3 of which stemming from PID. At Belle~II, it should be possible to match the BaBar systematics for all sources and---having an ultimately much larger data sample---to further reduce PID-related uncertainties.
While specific systematic studies need to be refined, preliminary estimates indicate that a reduction of the uncertainty on the HVP contribution to $g-2$ to 0.4\%, is within reach~\cite{Belle-II:2018jsg}. This would be comparable in size to the expected uncertainties on $a_\mu$ from the g-2 experiment.
Examples of planned improvements, facilitated by the large Belle~II data set, are the exploitation of the different distribution of the $\pi\pi$ channel compared to $\mu\mu$ as a result of the different spins of these particles. Improved QED Monte Carlo simulations are also expected to play a role.
It is worthwhile to point out that while the $\pi\pi$ channel is the dominant one, Belle~II will also be able to provide results on other contributions. Similar to the $\pi^+\pi^-$ channel, there is considerable tension between different data sets for the $K^+K^-$ channel. Furthermore, channels with more than two pions/kaons or channels involving $K_L$ are partially not well known and could be investigated at Belle~II.

\subsection{Constraining HVP through hadronic $\tau$ decay}

An alternative approach of reducing the uncertainty on $a_\mu^\textrm{HVP}$ is using hadronic data from semileptonic $\tau$ decays, useful to complement $e^+e^-$ data at low $s$ below about 1.5 GeV, and first employed in Ref.~\cite{Alemany:1997tn}.\footnote{For a more recent evaluation see, e.g., Ref.~\cite{Miranda:2020wdg}.} The invariant-mass spectra of the hadronic state $\mathcal{H}^-$ in $\tau \to \mathcal{H}^- \nu_\tau$ can be related to the corresponding isovector final state in $e^+e^- \to \mathcal{H}^0$. In particular, the  $e^+e^-\rightarrow \pi^+\pi^-$ process is related via so-called {\em conserved vector current} (CVC) relations to $\tau \to \pi^0 \pi- \nu_\tau$ decays. Semileptonic $\tau$ decays have been extensively studied, e.g., at LEP and the B-factories. 

\begin{figure}
    \centering
    \includegraphics[width=0.6\textwidth]{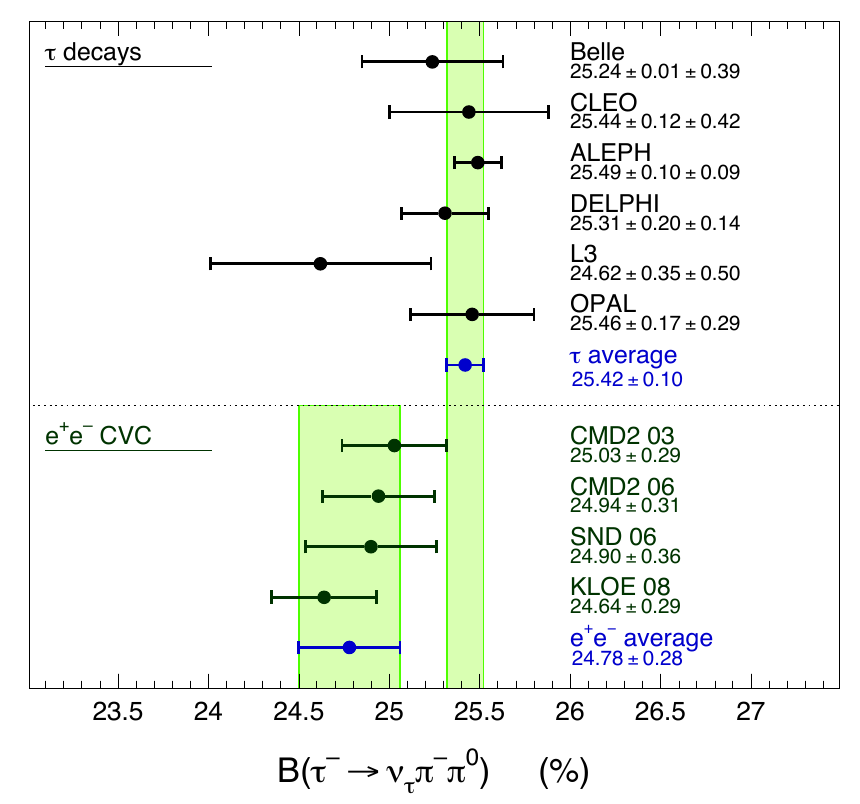} 
    \caption{Measured branching fractions for $\tau \to \pi^0 \pi- \nu_\tau$ compared to  predictions using $e^+e^-\rightarrow \pi^+\pi^-$ data, applying isospin-breaking corrections. (Figure from Ref.~\cite{Davier:2010fmf})}
    \label{fig:tau_vs_ee}
\end{figure} 

A disadvantage of using $\tau$-decay data is the need to account for isospin-breaking (IB) effects due to the mass difference between up and down quarks. From Fig.~\ref{fig:tau_vs_ee} it becomes clear that the use of data on $\tau$-decays does not currently help to reduce uncertainties on $a_\mu^\textrm{HVP}$ due to the tension with $e^+e^-\rightarrow \pi^+\pi^-$ data. Therefore, further investigation is clearly needed, in particular on the treatment of IB. The  data set of Belle~II will provide further input in this undertaking, and reduce uncertainties on the spectral functions of $\tau$ decays.


\subsection{Constraining HLbL via $e^+e^- \to e^+e^-h$ reactions}

With the HVP contributions at higher orders well under control, the next leading hadronic uncertainty comes from the HLbL contribution, for which the current estimate $a_{\mu}^{\textrm{HLbL}} = 92(19)$~\cite{Aoyama:2020ynm,melnikov:2003xd,masjuan:2017tvw,Colangelo:2017fiz,hoferichter:2018kwz,gerardin:2019vio,bijnens:2019ghy,colangelo:2019uex,pauk:2014rta,danilkin:2016hnh,jegerlehner:2017gek,knecht:2018sci,eichmann:2019bqf,roig:2019reh} is, by contrast to the HVP, in good agreement with lattice QCD results~\cite{Blum:2019ugy,Chao:2021tvp}. Still, the expected experimental uncertainty requires an improvement of the current uncertainties and to better understand certain aspects. 
Compared to the HVP, relating this contribution to data is much more challenging, yet a dispersive approach has been put forward recently that allows to relate unambiguously different hadronic contributions to the HLbL tensor~\cite{Aoyama:2020ynm}. In particular, the lowest-lying channels, the $\pi^0$, $\eta$, and $\eta'$ poles~\cite{masjuan:2017tvw,hoferichter:2018kwz,gerardin:2019vio}, as well as the $\pi\pi$ and $K\bar{K}$~\cite{Colangelo:2017fiz,eichmann:2019bqf,Danilkin:2021icn,Miramontes:2021exi} channels, have been estimated to the required precision. 

However, incorporating higher-multiplicity channels such as 3 hadrons in a purely dispersive framework
becomes a challenging task, as it is to deal with the high-energy regime due to the several scales involved~\cite{melnikov:2003xd,bijnens:2019ghy}. It is expected that such multi-hadron contributions are only relevant for $g-2$ in the resonant region, which can be safely approximated to the required precision by the corresponding resonances that should be embedded in the dispersive framework. In this context, one of the prominent roles is played by axial-vector meson resonances, which are comparatively less precisely determined than the pseudoscalar poles or $\pi\pi$ channels~\cite{Aoyama:2020ynm}. 
In particular, these have been recently revisited in Refs.~\cite{Cappiello:2019hwh,Leutgeb:2019gbz,Masjuan:2020jsf}, suggesting a substantial contribution ($\sim 15\times10^{-11}$) that could be twice as large as previous estimates~\cite{Aoyama:2020ynm}, mainly due to their high-energy behavior. Beyond that, they have been found to play a prominent role in fulfilling certain short-distance constraints~\cite{Cappiello:2019hwh,Leutgeb:2019gbz,Knecht:2020xyr,Masjuan:2020jsf} that are important to better understand the HLbL. Overall, axial-vector meson contributions definitely deserve further scrutiny.

To improve such contributions, a better knowledge of their form factors describing the $\gamma^*\gamma^* \to A$ transition is required, in particular for the lowest-lying multiplet $a_1(1260)$, $f_1(1285)$, $f_1(1420)$. Such a transition can be described in terms of 3 independent (1 symmetric, 2 antisymmetric) form factors that depend on the photons' virtualities~\cite{roig:2019reh,Hoferichter:2020lap,Zanke:2021wiq} (see also \cite{Rudenko:2017bel,Milstein:2019yvz}), while it is the symmetric one that plays the most prominent role at low energies as well as in short-distance constraints. 
At present, there is only limited information for the $f_1(1285)$, $f_1(1420)$ states from the L3 Collaboration \cite{L3:2001cyf,L3:2007obw} in the singly-virtual regime (i.e., one quasi-real photon, or $\gamma^*\gamma\to A$) for photon virtualities below $Q^2=6~\textrm{GeV}^2$. Additional information, also at higher invariant masses, would be valuable to further constrain them and to better understand their asymptotic behavior that, in contrast to the pseudoscalar mesons, is expected to behave as $Q^{-4}$~\cite{Hoferichter:2020lap}. 
Further, the contribution from $a_1(1260)$ is inferred via isospin symmetry from the available information on the $f_1(1285)$ and $f_1(1420)$ cases, which, however, introduces assumptions and requires prior knowledge of their mixing. It would be valuable to have experimental information about it. In particular, while due to charge factors the most relevant axial-vector meson contribution to $g-2$ is the $f_1(1285)$, the $a_1(1260)$ plays an important role in the hyperfine structure contribution in muonic hydrogen~\cite{Dorokhov:2017nzk,Miranda:2021lhb}. Belle~II could provide valuable information about this form factor. Some recent study regarding double-tagged measurements for the $f_1(1285)$ case can be found in~\cite{Szczurek:2020hpc}.

While the axial-vector meson form factors are gaining attention for the reasons described above and the few available data at disposal, there are further measurements that could be performed at Belle~II and are certainly interesting. Among them is the $\pi^0$ transition form factor, which can be accessed in the $e^+e^-\to e^+e^-\pi^0$ reaction, and which would be helpful to clarify the current tension among BaBar~\cite{BaBar:2009rrj} and Belle~\cite{Belle:2012wwz}. Likewise, the corresponding measurement for the $\eta$ and $\eta'$ would be useful as well, since only BaBar data is available at high energies, which is important to infer properties about their asymptotic behavior. Finally, $e^+e^-\to e^+e^-\pi^+\pi^-$ can be useful to further cross-check current approaches.

\section{Understanding hadronization}
\label{sec:hadronization}
The process of hadronization describes how final state hadrons, that can be detected, are formed from partons. Since it is governed by non-perturbative dynamics, it cannot be calculated analytically and, contrary to nucleon dynamics, up to now, the description of hadronization, has been elusive to lattice calculations~\cite{Metz:2016swz}.

Having an accurate description of hadronization is important for several fields of physics and a relevant physics program at Belle~II will be described in the subsections of this chapter. 

\begin{itemize}
    \item Hadronization is a fundamental process in non-perturbative QCD. Studying it, will shed light on QCD dynamics in hadron formation that are not accessible otherwise. One example is the creation of final state mass which will be discussed in section~\ref{sec:jet_mass}. 
    Other examples are tests of emerging calculations in jet production, see Sec.~\ref{sec:jets} and TMD physics probed in energy-energy calculations, see Sec~\ref{sec:eecorr}. Here, the clean $e^+e^-$ environment is very favorable. As will discussed more below, the statistics of Belle~II are needed to map out the full multi-dimensional dependencies of the hadronization process and the correlations that can shed light on the relevant dynamics.
    
    \item The traditional way to analytically compute cross-sections for (high-$p_\mathrm{T}$) hadron production uses parton-to-hadron fragmentation functions (FFs)~\cite{Metz:2016swz}. Fragmentation functions encapsulate the non-perturbative aspects of hadronization in a factorized pQCD calculation. As such, the knowledge of FFs is necessary to extract parton distribution functions, which in turn describe the quark-gluon dynamics inside the nucleon, from scattering experiments in a full QCD calculation.
    Precision measurements of fragmentation functions have been instrumental in extracting the spin averaged and spin-dependent nucleon structure~\cite{
    ANSELMINO2020103806}. Section~\ref{sec:hadronAndJet} will outline future needs, in particular by the planned EIC experiments, as well as planned contributions from Belle~II.
  The emphasis of the Belle~II program will be on the extraction of the full multi-dimensional dependency of fragmentation functions with complex final states. Examples for such final states are di-hadron correlations discussed in Sec~\ref{sec:dihadron} and polarized hyperons discussed in Sec~\ref{sec:lambda}.
  
  These final states are sensitive to spin-orbit correlations in hadronization their full kinematic dependencies have not been mapped out yet. However, they are important, as the additional degrees of freedom in the final states, allow a more targeted access to the nucleon structure in semi-inclusive deep inelastic scattering experiments (SIDIS), e.g., at JLab and the EIC. One recent example of this is the extraction of the twist-3 PDF $e(x)$ via di-hadron correlations, which is sensitive to the force gluons exert on a fragmenting quark as it traverses the nucleon remnant~\cite{Burkardt:2008ps,CLAS:2020igs,Hayward:2021psm,Courtoy:2022kca}.

    \item A detailed understanding of hadronization is important to model background and signal processes for new physics discoveries. Both at B-factories themselves, but also at the LHC. Currently, modeling of backgrounds originating from light quark fragmentation are mainly taken from Monte-Carlo Event Generators (MCEG), such as Pythia~\cite{Sjostrand:2014zea}, HERWIG~\cite{Bellm:2019zci} or Sherpa~\cite{Sherpa:2019gpd}. However, tuning those generators to a precision needed for discovery science requires the accurate reproduction of correlated hadro-production that can only be verified with clean SIA data. Currently, MCEGs mostly rely on LEP data. However, to have confidence in the extrapolation to LHC energies, it is essential to verify that those models also work at CM energies an order of magnitude below LEP.
     Furthermore, the hadronization of a large-mass system frequently
involves a natural subdivision into that of variable-mass smaller
systems, e.g., when a $g \to q\overline{q}$ parton-shower branching
splits one colour-singlet system into two.

A comprehensive program with the high statistics Belle~II data is also needed to  obtain the precision necessary for Belle~II analysis. Section~\ref{sec:mcTuning} outlines a program needed to tune MCEGs.
        MCEGs are also crucial for inference based models, e.g.~\cite{
    Brehmer:2019xox}, which will be more important in the future to 
    extract physical quantities. A recent development has been the extension of MCEGs to include spin orbit correlations. Sec.~\ref{sec:polmc} describes how Belle~II measurements can inform this novel spin-dependent MCEGs by benchmarking against spin dependent di-hadron correlations.
  
    \item Where MCEGs describe full events and FFs integrate over most of the event with the exception of the hadron in question (for single hadron FFs), more recently, intermediate observables describing more correlations in hadronization have gained more recognition in the field. 
    The di-hadron fragmentation functions mentioned above and discussed in Sec~\ref{sec:dihadron} are already an example using the language of fragmentation functions. Beyond current factorization theorems, there has been significant efforts recently to define correlation measurements that are sensitive to dynamics in hadronization and that can be interpreted in models (e.g., string models) and that while not yet realized, might be describable in a full QCD calculation with future, extended factorization theorems. These kind of correlation measurements have already been a focus at the LHC (see {\it e.g.}, the recent analysis in Ref~\cite{ATLAS:2020bbn}). At Belle~II precision correlation measurements using single hadrons can be made and Sec~\ref{sec:eecorr} introduces as an exemplary measurement correlations between leading particles. 
    
   \item An accurate knowledge of parton (in particular gluon~\cite{,dEnterria:2013sgr}) FFs into hadrons (both inclusively and for individual hadron species) in $\epem$ collisions is also of utmost importance to have an accurate ``QCD vacuum'' baseline to compare with the same objects measured in proton-nucleus and nucleus-nucleus collisions and thereby quantitatively understand final-state (``QCD medium'') modifications of the FFs~\cite{Albino:2008aa,Accardi:2009qv}.
\end{itemize}

The reaction of $e^+e^-$ annihilation has always been a method of choice to access hadronization in a clean environment. In $e^+e^-$ a $q\bar{q}$ pair is created in a known initial state, eliminating the need for the knowledge of other non-perturbative functions in the process. Therefore, Belle~II with its unprecedented dataset in spe, will be instrumental in exploring the topics summarized above and discussed in more detail below. For the first time, the integrated luminosity in an $e^+e^-$ experiment will be sufficient to explore the dynamics in hadronization through multi-correlation measurements.





\subsection{Hadron and jet fragmentation studies as input for the EIC}
\label{sec:hadronAndJet}
Electron-positron annihilation data are crucial to study the fragmentation of light quarks into hadrons. This process is described by FFs. For a general overview of this topic, see Refs~\cite{Metz:2016swz}. Since FFs are non-perturbative objects, they have to be measured in experiments. They can be seen as the time-like counterparts to parton distribution functions (PDFs). But unlike PDFs, they are currently inaccessible on the lattice. Similar to PDFs, the study of FFs can reveal aspects of QCD that are not directly evident from the Lagrangian.

However, FFs receive arguably most attention as necessary ingredients to extract aspects of the proton wave functions, e.g., encoded in PDFs, from semi-inclusive deep-inelastic scattering data where the fragmentation functions provide additional flavor and spin sensitivity.
For example, the first extraction of the distribution of transversely polarized quarks in a transversely polarized nucleon, the so called transversity PDF, which is one of the three collinear PDFs which are needed to describe the nucleon structure at leading twist, could only be extracted in a global fit including the first measurement of the transverse polarization dependent Collins FF at Belle~\cite{Belle:2005dmx,Belle:2008fdv}. The B-factories were the first $e^+e^-$ machines to record enough data to be sensitive to polarization and transverse momentum dependent FFs. These extractions had and still have a profound impact on the field of nuclear physics. 
The general case for the importance of the measurement of fragmentation functions for current and future programs to extract the nucleon structure from SIDIS and $pp$ experiments can be found e.g., in the whitepaper to the nuclear physics long range plan~\cite{eeOldWhitePaper}.

However, to interpret results from the current and next generation SIDIS experiments, such as JLab12 and the future Electron Ion Collider (EIC), which will collect orders of magnitude more statistics and have physics programs focusing on more sophisticated final states, such as polarized hyperons and hadron correlations, the statistics collected at the first generation B-factories will not be enough to achieve a fully multidimensional binning. It is well known, that  fully multidimensional binnings are needed to minimize model assumptions in the extraction of physics quantities. Inclusive results rely on model assumptions of the extracted physics for acceptance corrections and in global fits. As discussed before, statistics hungry final states with additional degrees of freedom, also provide for a more targeted access to properties of the nucleon structure. An example is the use of di-hadron correlations to extract the twist-3 FF $e(x)$ in SIDIS. Using traditional single-hadron measurements, the relevant observable contains a mix of several unknown PDFs and FFs~\cite{CLAS:2021opg}, whereas using di-hadron FFs, these ambiguity is under control~\cite{Courtoy:2022kca}

For light hadron spin and transverse momentum dependent fragmentation functions higher statistics would allow a more sophisticated multi-dimensional analysis which may be even of interest as input for transverse momentum dependent distributions functions relevant at the LHC. For heavier final states, such as hyperons or charmed bayrons, the fragmentation measurements have just begun at Belle \cite{Belle:2017caf} but statistics are in some cases still limited. There is a clear need for Belle~II to collect a data-set orders of magnitude larger to extract FFs for more luminosity hungry final states and to map them out in multiple dimensions. This need is even more urgent as the nuclear physics community just committed to build the EIC at Brookhaven National Laboratory over the next decade at a cost which is currently projected to be between $\$1.6-\$2.6$ billion. It is also envisioned to operate Belle~II with a polarized $e^-$ beam from 2026 onwards, collecting between $20-40~\textrm{fb}^{-1}$ in that configuration. This would enable the measurement of a new class of observables sensitive to the hadronic mass generation~\cite{Accardi:2017pmi,Accardi:2019luo}, which is described in Sec.~\ref{sec:jet_mass}.
Compared to Belle, measurements of fragmentation functions (or hadronization in general), will also profit from the upgrades to the detector, in particular particle identification and vertexing~\cite{Belle-II:2018jsg}. The improved vertex resolution will help to discriminate against charm production. Using the vertex will likely have less bias then using D tagging as has been done in Belle analyses, see e.g. Refs.~\cite{Belle:2005dmx,Belle:2008fdv}.

Here we will describe two exemplary channels, which will be enabled at Belle~II, measuring di-hadron correlations fully differential in Sec.~\ref{sec:dihadron} and a polarized $\Lambda$ program in Sec.~\ref{sec:lambda}. 
In addition to semi-inclusive hadron production, jet physics will play an important role in accessing the three-dimensional nucleon structure at the EIC~\cite{Boughezal:2018azh,
PhysRevD.102.074015}. The corresponding theory is very much an active field of development and data from Belle~II would provide ideal precision tests for the framework~\cite{Gutierrez-Reyes:2019msa,Gutierrez-Reyes:2019vbx}. In particular the addition of jet substructure measurements may help to increase the flavor sensitivity \cite{Fraser:2018ieu} at an EIC as described in Sec.~\ref{sec:jets}.

\subsubsection{Studies of hyperon production}
\label{sec:lambda}
Hyperon physics will be one of the headline measurements in hadronization measurements at Belle~II. 
Due their self-analyzing decays, the production of hyperons such as $\Lambda's$ opens up the possibility to study spin orbit correlations in hadronization~\cite{Pitonyak:2013dsu}. 
This in turn gives sensitivity to fundamental properties of QCD that also have an analogy in the nucleon structure studies such as tests of modified universality~\cite{Boer:2010ya}, which is one of the milestones of current nuclear physics research~\cite{Aprahamian:2015qub} since it is a non-trivial consequence of the gauge structure of QCD~\cite{Collins:2002kn, Brodsky:2013oya}
Additionally, there are some longstanding puzzles in hyperon related observables that can only be answered by precision measurements in $e^+e^-$. One example is the observation of large transverse polarizations in $\Lambda$ production with unpolarized beams~\cite{Bunce:1976yb,Heller:1996pg}, a phenomenon that is still not completely understood and that has been one of the motivating factors of a large program of spin structure measurements at several facilities over the last few decades. 
Consequently, the first measurement by Belle~\cite{Belle:2018ttu} prompted strong theoretical activity~\cite{Gamberg:2018fwy,Anselmino:2019cqd,DAlesio:2020wjq,Boglione:2020cwn,Callos:2020qtu,Gamberg:2021iat,Chen:2021zrr,Li:2020oto} it will also be the base for a hyperon program at the EIC~\cite{Kang:2021kpt}. 

\begin{figure}[b]
    \centering
    \includegraphics[width=0.55\textwidth]{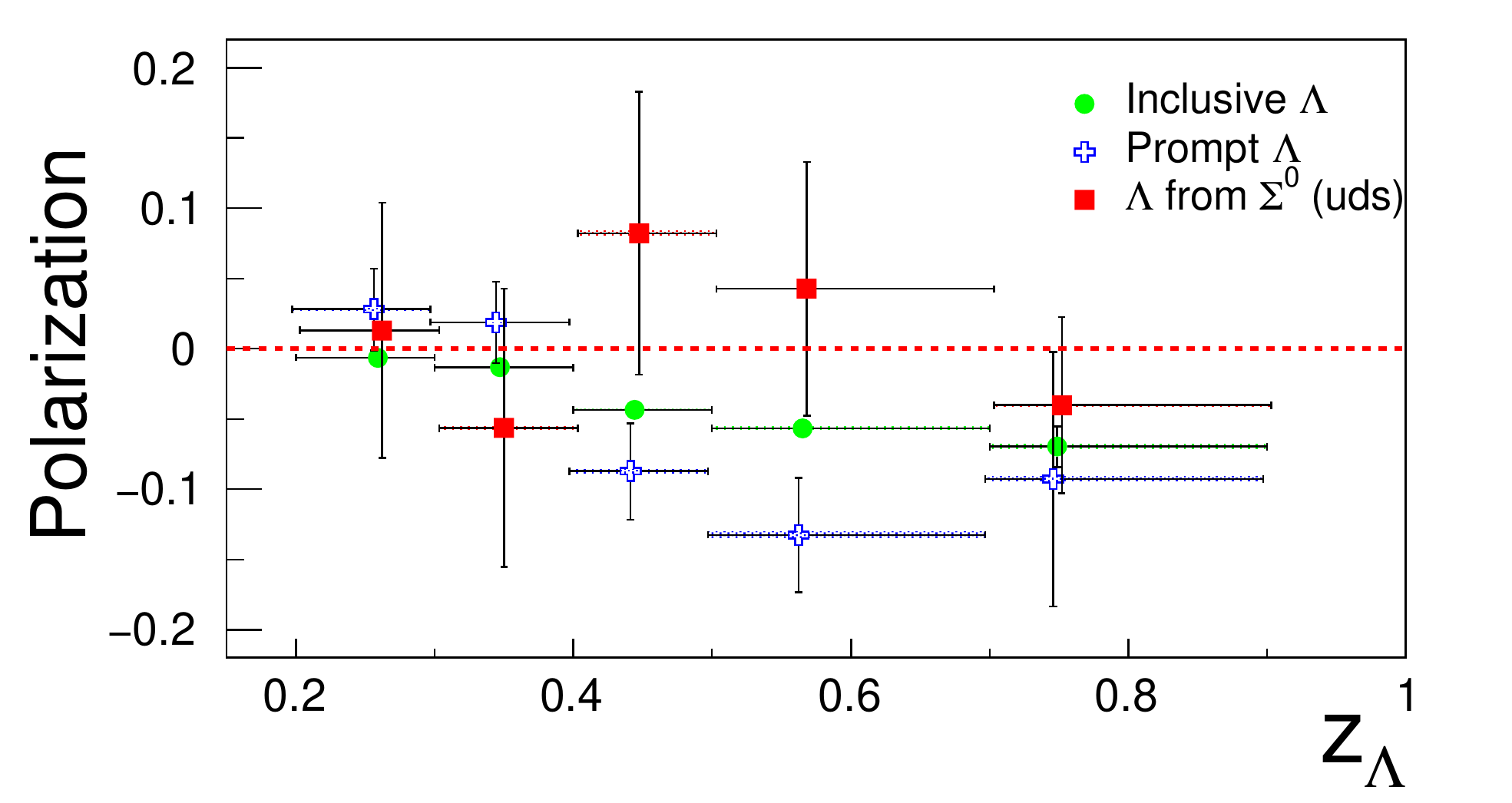}
    \caption{Transverse $\Lambda$ polarization in inclusive production in $e^+e^-$ annihilation for inclusive and prompt $\Lambda$'s, as well as from $\Sigma^0$ decays (for details, please see Ref.~\cite{Belle:2018ttu}). The need for a higher statistics dataset is evident. Figure taken from Ref.~\cite{Belle:2018ttu}.}
    \label{fig:unfoldedLambdaResults}
\end{figure}

Figure~\ref{fig:unfoldedLambdaResults} shows the previous results by Belle \cite{Belle:2018ttu} for the transverse polarization of $\Lambda$ hyperons corrected for feed-down. An inclusive measurement is already statistically challenging, in particular when considering a multidimensional analysis in the relevant variables $z$ and $p_T$. However, for a clean theoretical interpretation, the $\Lambda$ signal has to be corrected for feed-down from heavier hyperons. In the Belle data, about 20\% of $\Lambda$'s produced in $uds$ production come from feed-down. An additional 30\% come from charm production, mainly via $\Lambda_c$ decays~\cite{Belle:2018ttu}. Due to the limited efficiency reconstructing the feed-down channels and the associated uncertainties, the resulting statistical uncertainties on the $\Lambda$ polarization are large using only the Belle data set as shown in Fig.~\ref{fig:unfoldedLambdaResults}. Using the final planned Belle~II data set would reduce the uncertainties on the asymmetries from the feed-down and the prompt $\Lambda$ to a level where it is small compared to the expected asymmetries, e.g., for a 50 times increase in luminosity, the resulting statistical uncertainties with the same binning as used in the existing Belle analysis~\cite{Belle:2018ttu} shown in Fig.~\ref{fig:unfoldedLambdaResults} can be reduced to below 1\%.
It should be mentioned that so far only the transverse $\Lambda$ polarization in the TMD picture, {\it i.e.} using the thrust axis or the axis of a hadron from associated production. It will be important to also measure the $\Lambda^\uparrow$ polarization in the twist-3 picture, {\it i.e.} with respect to the plane spanned by beam axis and $\Lambda$ momentum~\cite{Gamberg:2018fwy}.


\subsubsection{Di-hadron production}
\label{sec:dihadron}
\begin{figure}
    \centering
   \includegraphics[width=0.9\textwidth]{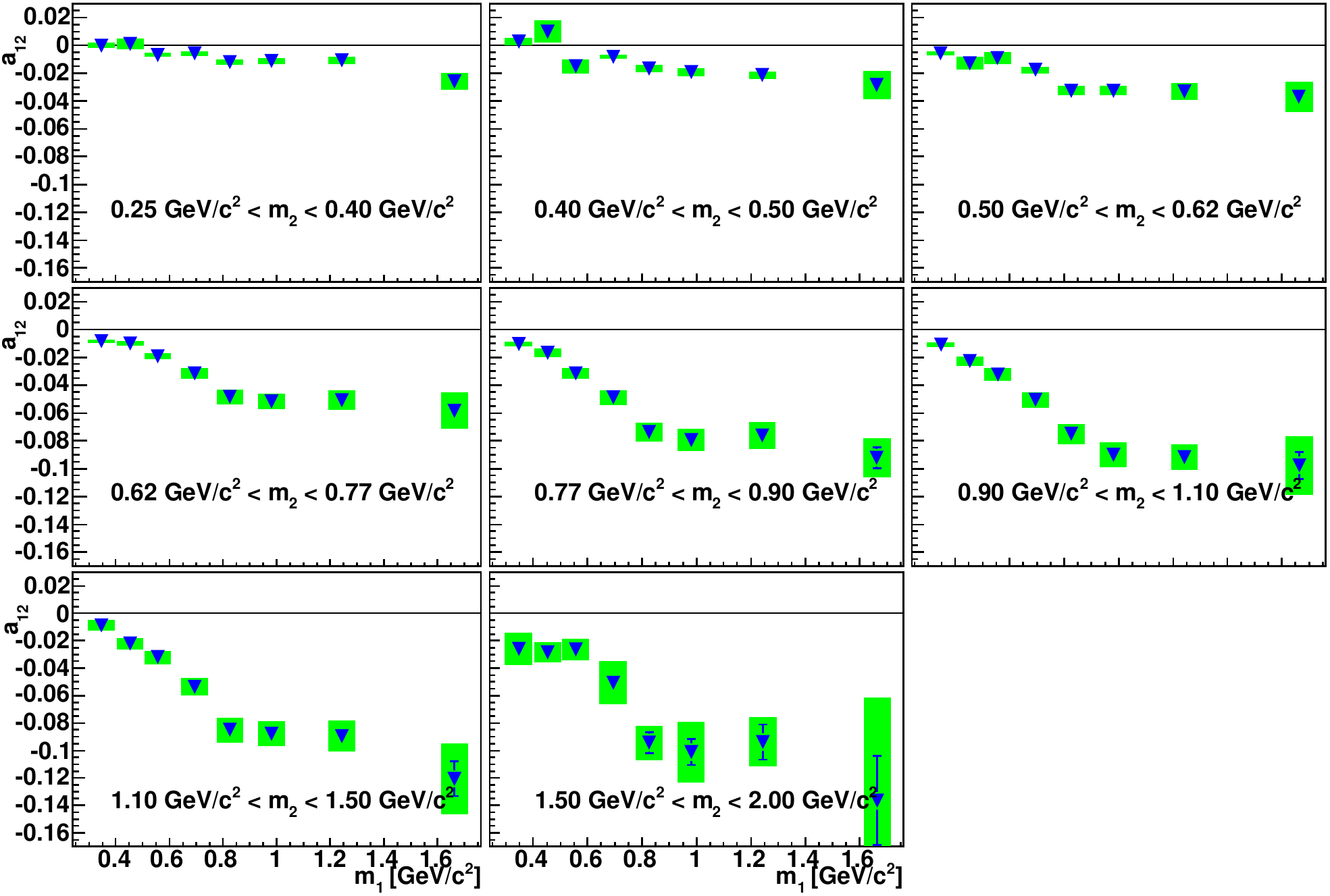}
    \caption{Transverse spin dependent IFF asymmetries binned in $m_1$, $m_2$ where the $m_i$ are the invariant mass of the hadron pair in each hemisphere. The full Belle dataset is used. While the uncertainties are generally small, high invariant mass bins show non-negligible statistical and systematic uncertainties. For a fully differential binning, in particular in $\theta$ and if the transverse momentum dependence is to be taken into account, more data would be needed. Figure from Ref.~\cite{Belle:2011cur}}
    \label{fig:diHadIff}
\end{figure}

\begin{figure}
    \centering
    \includegraphics[width=0.98\textwidth]{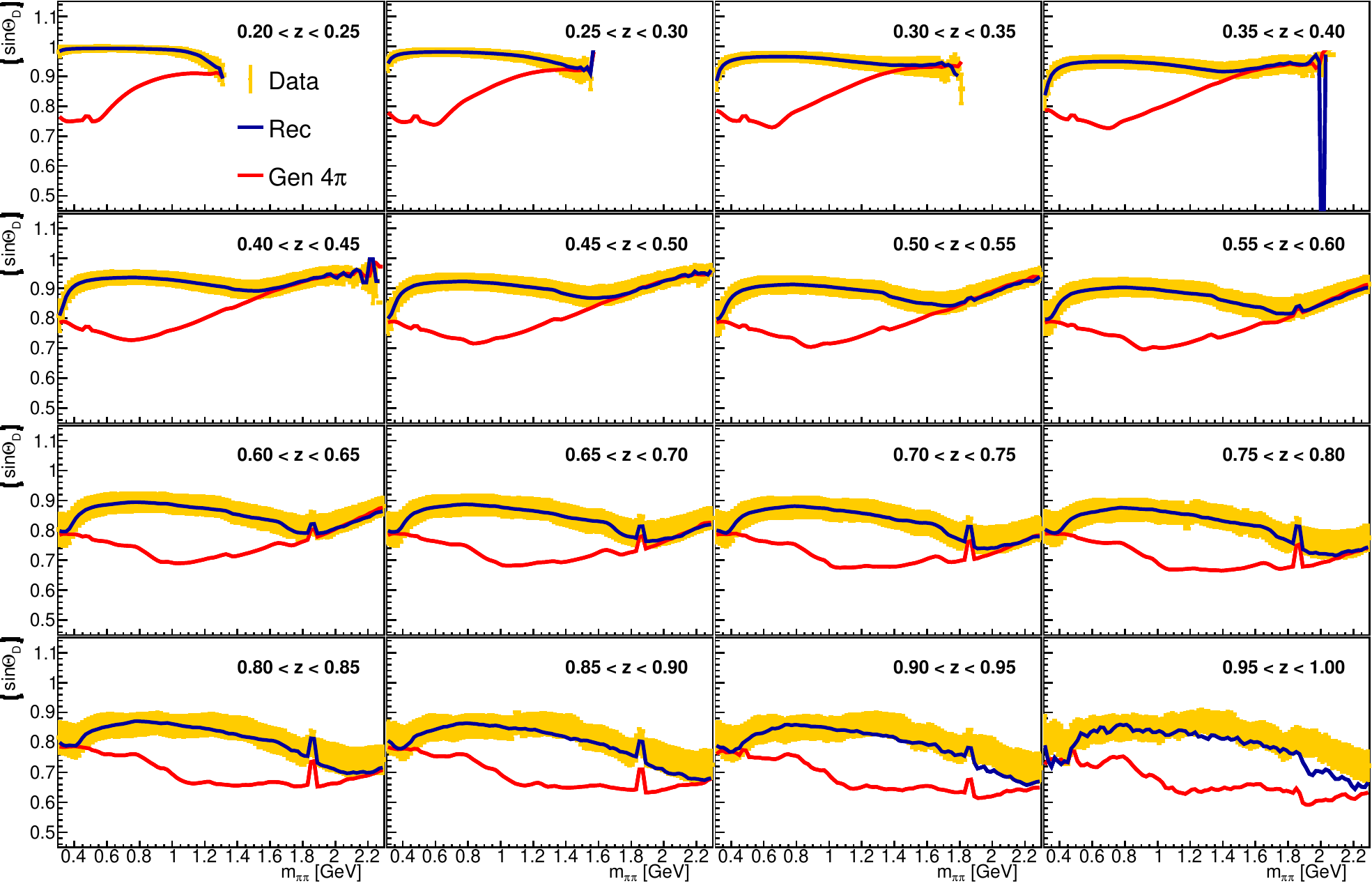}
    \caption{The $\sin(\theta)$ moment, which is a common to all partial waves, extracted from data compared to simulations with and without acceptance. There is a significant acceptance effect which might lead to a model dependence of the fragmentation function extraction because the mix of partial waves is not precisely determined. Figure taken from~\cite{Belle:2017rwm}}
    \label{fig:diHadThetaDependence}
\end{figure}

As opposed to inclusive single-hadron fragmentation functions, di-hadron correlations capture the dynamics in the hadronization process. 
In a factorized picture, di-hadron fragmentation is described by di-hadron fragmentation functions~\cite{Metz:2016swz, Pisano:2015wnq}. These functions are in the collinear case dependent on $z$ and the invariant mass $m$ of the hadron pair. In the TMD case, they also acquire transverse momentum dependence. They can be developed in a partial wave expansion, which gives additional dependence on angles $\theta$ and $\phi$. The $\phi$ dependence can be used to access, e.g., the dependence on the longitudinal or transverse polarization of the fragmenting quark, whereas the combination with the $\theta$ dependence gives the complete description of the correlation. If the $\theta$, $\phi$ dependencies are interpreted as partial waves, additional insights on the QCD amplitude level can be gained, as the various partial waves can be interpreted as interference terms between amplitudes with different quantum numbers. For example, early model calculations assumed that the dominant contribution to the important transverse polarization dependent DiFF $H_1^\sphericalangle$ into charged pion pairs originates from the interference of $s-$ and $p-$ wave components, with the former coming from non-resonant pion production and the later coming from unpolarized $\rho$ decays.
However, recent preliminary results from CLAS12~\cite{Dilks:2021nry} indicate that the real mechanism is likely more complex.
 More details on di-hadron FFs can be found in Ref~\cite{Pisano:2015wnq,Metz:2016swz}.
From this brief description it is already clear, that the di-hadron correlations are very interesting to characterize the dynamics in hadronization. They can for instance be compared with newly developed polarized Monte Carlo physics simulations, as discussed in Sec.~\ref{sec:polmc} and, given their additional degrees of freedom, they can be used to access the nucleon structure in hard scattering experience in a more targeted manner as was discussed in the beginning of this document.
Due to the dependence on $z$, $m$, $p_t$, $\phi$ and $\theta$, a complete measurement is quite statistics hungry. Even more though, as for chiral-odd FFs, such as the transverse spin dependent ones, back-to-back correlations of di-hadron pairs have to be measured. Belle did a measurement of transverse spin dependent effects in di-hadron correlation~\cite{Belle:2011cur} and the di-hadron cross-section~\cite{Belle:2015hut,Belle:2017rwm} integrated over $p_t$ and $\theta$ dependence. 
Figure~\ref{fig:diHadIff} shows the Belle results for the spin dependent results~\cite{Belle:2011cur}. It can be seen that for high $m$ and $z$ bins the statistical power is starting to lack. This indicates that a fully multidimensional binning including the $\theta$ or $p_T$ dependence needs the statistical power of Belle~II. Figure~\ref{fig:diHadThetaDependence} shows the $\sin(\theta)$ moment from the measured cross-section for generated and reconstructed simulations as well as for data. It is clear that there are significant acceptance effect for this $\theta$ dependent factor which is common to all partial waves. Therefore, a measurement differential in $\theta$ is important not only for the partial wave decomposition, but also so that the eventual results do not contain a complicated mix of partial waves. As already discussed earlier in this section, recent results from CLAS12 indicate that, contrary to initial expectations, for the transverse polarization dependent DiFFs $H_1^\sphericalangle$ and $G_1^\perp$, many partial waves contribute to the signal. This is important as $H_1^\sphericalangle$ so far has attracted the most attention, due to its sensitivity to the quark transverse polarization and it stands to reason that this is a finding that holds true for other di-hadron FFs as well, in particular at higher energies such as the EIC as more channels open up. The FF $G_1^\perp$ is sensitive to the longitudinal quark polarization.


\subsection{Interrelation with MC models}\label{sec:tuneMC}
As outlined above, the precise knowledge of fragmentation functions extracted from $e^+e^-$ annihilation is essential for precision physics at the EIC. As discussed below, multidimensional measurements as well as the measurements of more complex final states, will profit significantly from Belle~II's additional statistics. However, even measurements that are currently systematic limited, like the Collins asymmetries for pions or hadron cross-sections {\bf will profit from a comprehensive QCD program at Belle~II}. 
Currently, the systematic uncertainties are dominated by MC uncertainties, in particular the differences between different tunes. This leads to systematic uncertainties for example of acceptance effects or the impact of initial state radiation on observables. The larger data set offered by Belle~II will open opportunities to significantly reduce these uncertainties in two different ways. Firstly and most importantly, a comprehensive program in collaboration with MCEG groups to tune hadronization models as outlined Sec.~\ref{sec:mcTuning} will significantly reduce uncertainties related to the MC model as it will allow to reduce the reasonable parameter space in these models. This will obviously also be important to other physics channels at Belle~II and other experiments.
Secondly, in multidimensional hadronization measurements, MC statistics limit the precision with which systematic uncertainties can be determined. For instance, for the di-hadron measurement shown in Fig.~\ref{fig:diHadIff} the systematic and total uncertainties in the highest invariant mass bins are dominated by the limited MC statistics.
Even though, reducing these uncertainties are not usually associated with the need for an entire experimental program, the generation of a sufficient amount of simulated, newly tuned MC, will require significant resources that can only be marshalled by a new collaborative effort over a extended time frame. For comparison, for Belle, six times of the original data is available for simulated events for light quark pair creation and ten times the original data for $\Upsilon(5S)$ production, but for only one set of MC parameters. To significantly improve on those statistics, a production of the planned MC size at Belle~II is needed. Belle~II plans to produce twice the amount of simulated data than data up a integrated data luminosity of 5~ab$^{-1}$ and an equal amount after that. It is estimated that a MC production of this size needs several PB of disk and tape space, as well as several hundred kHEPSpec06 CPU resources. Clearly, only a large collaboration can provide these resources.

\subsubsection{Tuning hadronization models in Monte Carlo event generators}
\label{sec:mcTuning}

It is general consensus that Monte Carlo event generators (MCEGs) are an essential part of modern high-energy physics.  
They are not only of practical importance to describe the underlying parton-level events in hadronic collisions,  but as they become more sophisticated actually provide tools to understand the complex dynamics in hadronization via physics based models. For details on role of event generators in high-energy physics, see the separate Snowmass contribution given in Ref.~\cite{Campbell:2022qmc}.
For many aspects of discovery physics, MCEGs have replaced analytical calculation and it is to be expected that the same might happen in nucleon structure physics as MCEG based inference methods become more popular. However, a major drawback of MCEGs is the need to tune model parameters to data. In order to reach the precision needed by current and future analyses, the information from the clean environment of $e^+e^-$ annihilation is indispensable. 
The latest generation of MCEGs, Pythia~\cite{Sjostrand:2014zea}, SHERPA~\cite{Sherpa:2019gpd} and HERWIG~\cite{Bellm:2019zci}, relies mainly on LEP data to describe hadronization~\cite{Buckley:2009bj}. Data from the B-factories have not been extensively used, due to the difficulties of incorporating data at lower center-of-mass energies. However, the consensus in the field is that theory progress enables the inclusion of B-factory data~\cite{energyDepMC} and that these data are in fact necessary to gain confidence in the energy dependence description and the predictive power of MCEGs. 
This is on one hand due to the superior statistics (Belle, e.g., collected about 1000 times the integrated luminosity of the LEP experiments, but on the other hand, together with the LEP data, B-factory data provides a lever-arm in center-of-mass energies that can give confidence for an extrapolation to LHC energies. Without a comprehensive program at B-factories, ideally at Belle~II, there is no confidence for the energy extrapolation. 
It should be noted though, that hadronization in $pp$, in particular at lower scales, can involve additional complex dynamics that are not present in $e^+e^-$ annihilation and not reflected in FFs. One example are color reconnections to the beam remnant. Therefore, the $e^+e^-$ data will be important to study many aspects of the hadronization model but certainly has to be supplemented with additional hadronic studies.

Such a program at Belle~II would primarily be based on the program characterizing hadronization using LEP data, which inform the following preliminary list of planned measurements. 


As a baseline, similar measurements that have been performed by LEP experiments and that have been important inputs to MCEG model tunings have to be performed. 
A incomplete list of measurements is given below. The measurements focus on correlations sensitive to the employed hadronization mechanisms.
As discussed several times in this text, correlations are useful in general to learn about dynamics, therefore other correlation measurements, e.g. of di-hadrons discussed in Sec.~\ref{sec:dihadron}, will also be important to benchmark MC. In particular, the comparison between quark polarization dependent correlations with polarization dependent MC.
This has not been done with LEP data, since the polarization dependent MCEG only became available recently. Section~\ref{sec:polmc} gives a brief description of the planned comparison with polarized MC.

Ideally, the measurements of the quantities for the non-polarized MCEG described below are performed separately for light quarks (u,d,s), charm tagged continuum, as well as $\Upsilon$ production. 

An initial set of measurements should constrain the following quantities:


\begin{itemize}
\item ``Traditional'' event shapes like thrust and (linear) sphericity.
\item Jet rates as a function of resolution, hemisphere masses and other pertinent variables.
    \item Identified particle spectra including $\pi, K, p, \gamma/\pi^0$.
    Belle already published particle spectra for charged mesons, protons and heavier baryons~\cite{Belle:2020pvy,Belle:2017caf} differential in $z$, the fractional momentum of the parent quark carried by the detected hadron, as well as vs. the transverse momentum with respect to the thrust axis~\cite{Belle:2019ywy}.
    However, be more sensitive to the event topologies, these results need to be reproduced differential in the relative momenta with respect to the event plane, $p_L, p_{\perp}$ and $p_{\perp\mathrm{out}}$.
    \item Multiplicities of resonance production in particular $\rho$, $\omega$ $K^*$, $\phi$, $\Lambda$ $\Sigma$, $\Xi$,  $\Omega$ as well as L\,=\,1 mesons. Especially the production ratios between pseudo-scalar and vector mesons are of importance, not only for the description of nuclear- and particle-physics events but, e.g., also for ultra-high-energetic cosmic air showers~\cite{dEnterria:2011twh,Sjostrand:2021dal}. 
    \item 
    Charge/strangeness/baryon number compensation along the event axis for two-jet topologies.
    Specific measurements needed are
    \begin{itemize}
    \item net compensation between positive and negative charges along the event axis in bins of rapidity.
    \item net strangeness compensation for baryons along the event axis in bins of rapidity. Here, in addition to multiplicities, a characterization of the particle production process is desirable.
    \end{itemize}
    \item Rapidity-ordered particle chains for exclusive final states. Those should be reconstructed back to resonances as much as possible. A comparison of the rates of events with with the same net particle content but different orderings should be made
 
\end{itemize}

It is also  noted, that for this comprehensive program, the integrated approach with theorists is important. In particular, an integration of future but also of legacy analysis in a framework like Rivet~\cite{Bierlich:2019rhm} which enables the seamless integration into MCEG bench-marking is important. This implementation is the basis for current automatic tuning methods~\cite{Buckley:2009bj}.

\subsubsection{Comparing to polarized MC models}
\label{sec:polmc}
\begin{figure}
    \centering
    \includegraphics[width=0.69\textwidth]{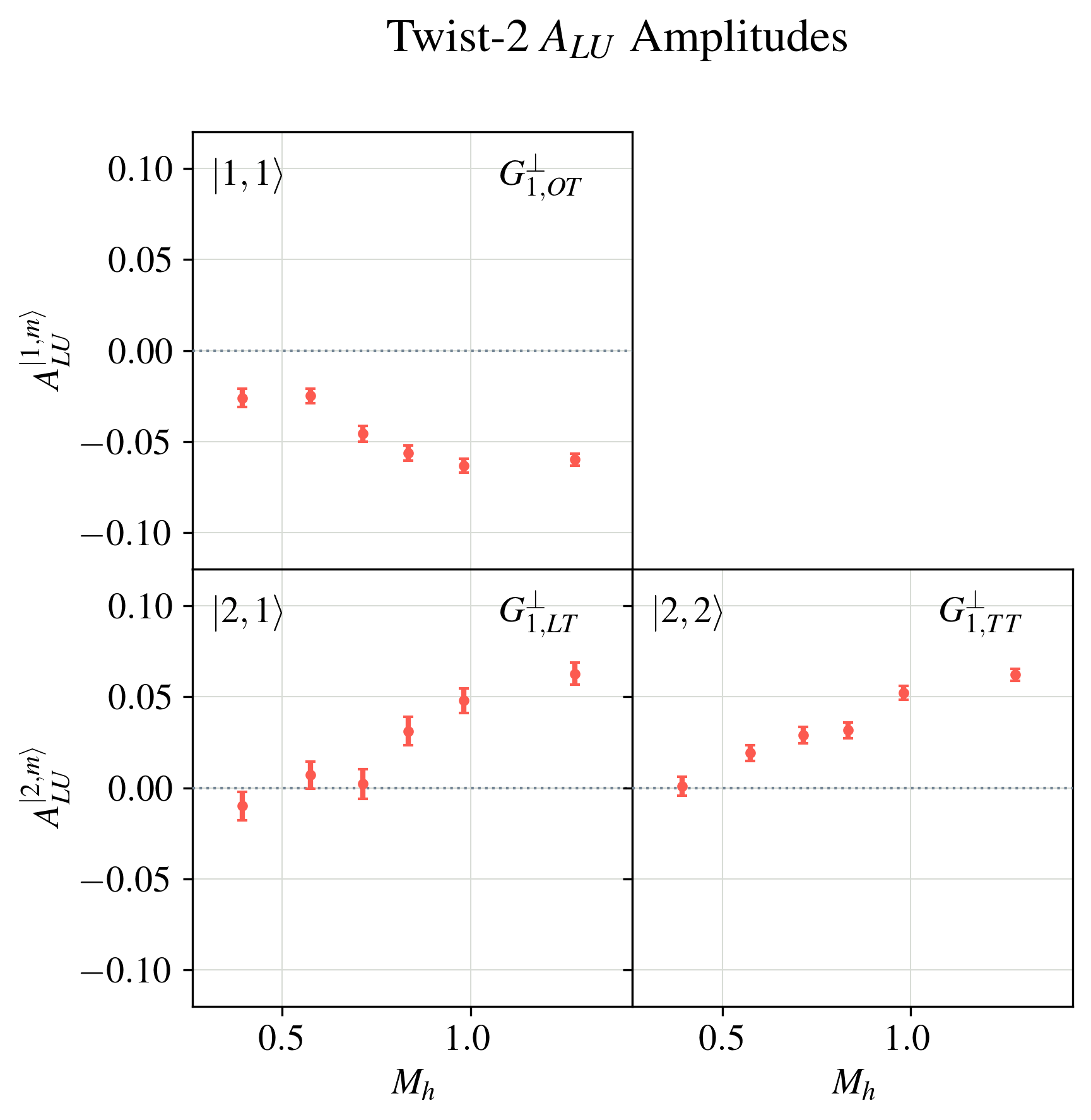}
    \caption{Beam spin asymmetries of di-hadrons at leading twist ($A_{LU}$) plotted versus the invariant mass of the di-hadron pair and decomposed into partial waves generated with the stringspinner model~\cite{Kerbizi:2021pzn} separated by partial waves up to L\,=\,2. The specific partial wave is indicated in the panel. Currently, the results show a similar trend but differ in some significant aspects from the observation in data shown in Fig~\ref{fig:dataPW} and indicate the importance of mechanisms not yet considered in the simulation, like the production of vector mesons. 
    \label{fig:stringSpinnerPW}}
\end{figure}

\begin{figure}
    \centering
    \includegraphics[width=0.69\textwidth]{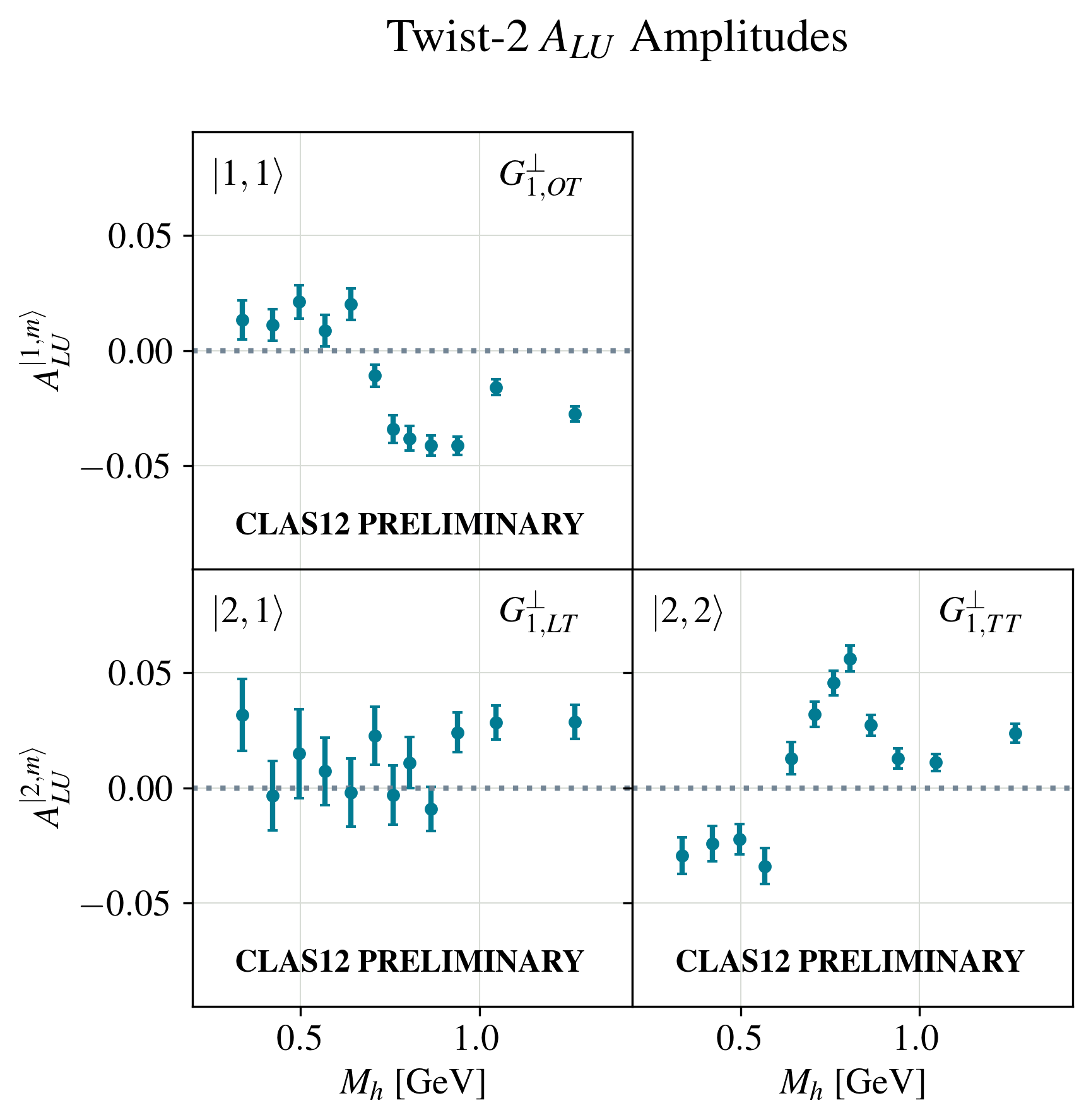}
    \caption{Beam spin asymmetries of di-hadrons at twist-3 ($A_{LU}$) plotted versus the invariant mass of the di-hadron pair and decomposed into partial waves up to L\,=\,2 as measured at CLAS12~\cite{Dilks:2021nry}. Significant differences in the shape of the observed asymmetries can be seen. For some partial waves, a resonance structure around the mass of the $\rho$-meson is visible. Figure from Ref.~\cite{Dilks:2021nry}}.
    \label{fig:dataPW}
\end{figure}

Spin dependent fragmentation is still relatively poorly measured and only recently has been included into stand-alone MCEGs or as additional packages for existing MCEGs \cite{Kerbizi:2018qpp}.
Modeling polarization dependent hadron production requires additional model assumption about the spin dependent dynamics in hadronization. Therefore bench-marking these generators will provide a first test of our understanding of these dynamics. 
At Belle~II, back-to-back hadron correlations, sensitive to the transverse polarization of the initial quark (Collins effect) as well as di-hadron correlations~\cite{Metz:2016swz, Belle:2011cur} would provide the needed benchmark processes. Even though, initial measurements are available from Belle~\cite{Belle:2005dmx,Belle:2008fdv, Belle:2011cur}, these have to be expanded to probe the polarization dependent dynamics in detail. For example any final state other than pions is still poorly measured and the feed-down from spin-1 mesons to pions and kaons appears to play a significant role in these polarized fragmentation models \cite{Kerbizi:2021gos}.
As a first step, precise measurements of single- and di-hadron correlations with final state kaons need to be made.
The feed-down contribution should be studied via the explicit measurement of either vector meson correlations or the comparison of measurements where pions and kaons are produced in a vector meson decay. Given the size of most vector meson resonances and the amount of  nonresonant background, $\phi$ decays may be of particular interest. However, for an interpretation the corresponding kaon Collins fragmentation function has to be measured with sufficient precision first  
In addition to existing measurements, which were only differential in one or two variables, capturing the complete dynamics requires the extension to multi-differential measurements. Here, the superior statistics of Belle~II will be required. 
One example is the partial wave decomposition of the di-hadron fragmentation function also discussed in Sec.~\ref{sec:dihadron}. Comparing these with models will provide important insights. As an example, Fig.~\ref{fig:stringSpinnerPW} shows the partial wave projections of the stringspinner model~\cite{Kerbizi:2021pzn} , which describes the $z$ and invariant mass dependence of the di-hadron fragmentation well~\cite{Kerbizi:2018qpp,Kerbizi:2021gos}.
In principle this can be compared with the partial wave decomposition of di-hadron correlations from CLAS12, shown in Fig.~\ref{fig:dataPW}, which are sensitive to the same fragmentation function coupled to the PDF $e$ or $f$. However, the stringspinner model does not include vector meson production yet which, as described above, are likely to play an important part in the polarized hadronization dynamics. Nevertheless, this comparison shows an interesting path forward. Belle~II data is important due to its high precision and independence from the nucleon structure. = 

\subsection{Jet physics and TMDs}
\label{sec:jets}

At high-energy collider experiments, energetic sprays of particles called jets can be observed in the detectors. Jets are emergent phenomena of QCD and exhibit a close relation to the underlying quark and gluon degrees of freedom. When quarks and gluons are produced at high energy, they emit radiation nearly collinear relative to their initial direction. By analyzing the jet production rate, jet correlations and their substructure, unique insights into QCD dynamics at the level of quarks and gluons can be obtained. At the LHC and RHIC, jets have been used frequently to test the Standard Model and to search for new physics. However, due to the difficult environment at proton-proton colliders, where a lot of soft radiation is produced due to the underlying event, low-energy aspects of jets are not well controlled and often removed systematically using so-called jet grooming algorithms~\cite{Dasgupta:2013ihk,Larkoski:2014wba}. The Belle~II experiment provides a clean environment to study jets at low energies, which is complementary to proton-proton collisions.
In addition, jets are expected to play an important role at the future EIC. In this regard, Belle~II can provide important information by studying universality aspects and constrain non-perturbative components. In the following, we review several possible future jet measurements at Belle~II. 
\begin{figure}[t]
\centering
\includegraphics[width=0.99\textwidth]{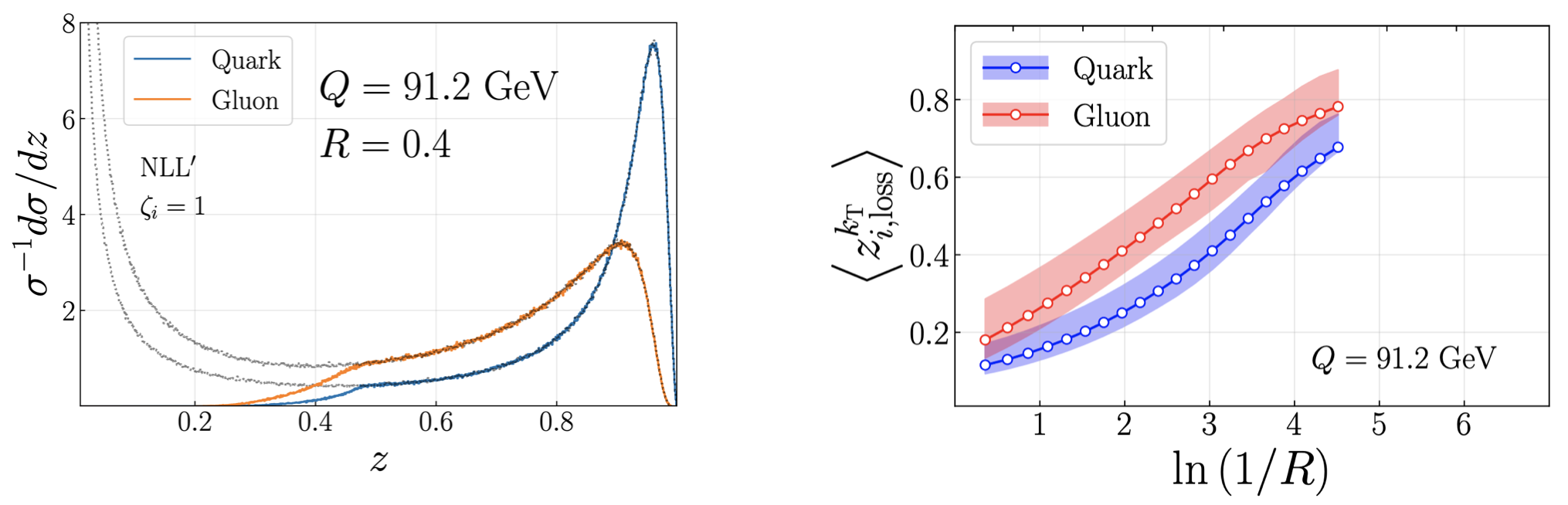}
\caption{Left: Numerical results for the inclusive and leading jet spectrum (single hemisphere) in $e^+e^-$ collisions. Right: The average energy loss of leading (hemisphere) jets. Figure taken from Ref.~\cite{Neill:2021std}.}
\label{fig:leading_jet}
\end{figure}

Different than inclusive jet production, leading and subleading jets are sensitive to the entire shower evolution and probe non-linear QCD dynamics. Recently, significant progress has been made in improving our understanding of the QCD factorization structure and evolution equations relevant for leading jet cross sections~\cite{Dasgupta:2014yra,Scott:2019wlk,Neill:2021std}. The factorization structure has to be modified as higher perturbative accuracy is achieved and jet functions that describe the transitions from quarks and gluons satisfy non-linear DGLAP-type evolution equations. Different than for inclusive jets, one has to distinguish between leading jets in one hemisphere and in the entire event since at least two jets are produced. Theoretical predictions for leading jet cross sections in $e^+e^-$ collisions at next-to-leading logarithmic (NLL$'$) accuracy at LEP energies are shown in the left panel of Fig.~\ref{fig:leading_jet}. In fact, recently LEP performed the first leading di-jet measurement~\cite{Chen:2021uws}. Also the ALICE Collaboration performed corresponding jet substructure measurements~\cite{Mulligan:2021ooe}. Similar experiments at Belle~II would allow for an extension of these measurements to low energies. In addition, the high statistics would allow for detailed measurements in the threshold region, where the leading jet energy close to half of the total center of mass energy. This region exhibits a strong sensitivity to threshold corrections and universal non-perturbative effects. Leading jet cross sections can be considered as normalized probability distributions to find a leading jet that originates for a quark or gluon. The radiation that is not contained in the leading jet is a measure of the jet energy loss, which plays an important role in heavy-ion collisions and in electron-nucleus collisions at the future EIC~\cite{Accardi:2009qv}. Numerical results for the (vacuum) energy loss of jets in $e^+e^-$ collisions are shown in the right panel of Fig.~\ref{fig:leading_jet}. Measurements at Belle~II would provide a clean baseline of vacuum energy loss and constrain its non-perturbative contribution. Similar measurements can be performed for hadrons, which would allow for the determination of the currently unconstrained (sub)leading hadron fragmentation functions. Detailed measurements of jet cross sections with different jet radii may also constrain the transition between jets and hadrons. Cross sections involving hadrons can be obtained from jet cross sections by formally taking the limit of a vanishing jet radius. Recently, it was possible to establish a theoretical formalism to quantitatively explore the transition from jets to hadrons using the parton-hadron duality hypothesis~\cite{Neill:2020tzl}. These results may lead to a new understanding the QCD fragmentation process and probe the boundary of perturbative QCD calculations.

\begin{figure}[t]
\centering
\includegraphics[width=0.99\textwidth]{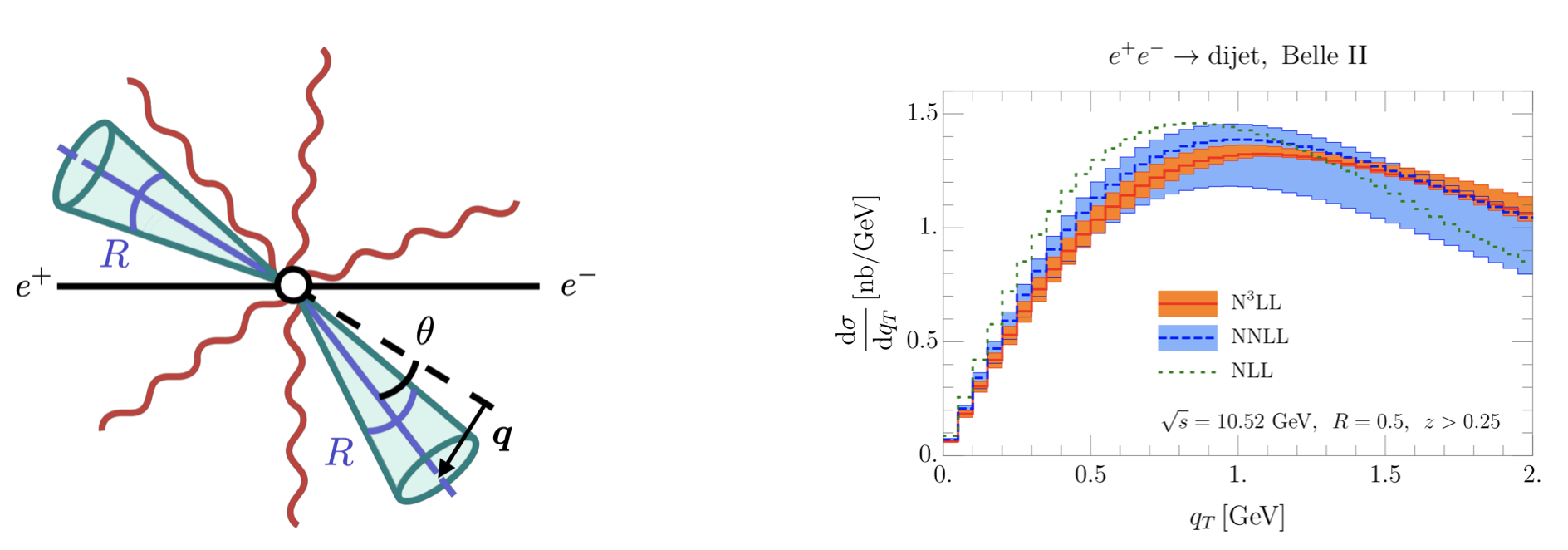}
\caption{Illustration and numerical predictions for di-jet correlations in $e^+e^-$ collisions at Belle~II. Figure taken from Ref.~\cite{Gutierrez-Reyes:2019vbx}.~\label{fig:dijet}}
\end{figure}

\begin{figure}[t]
\centering
\includegraphics[width=0.8\textwidth]{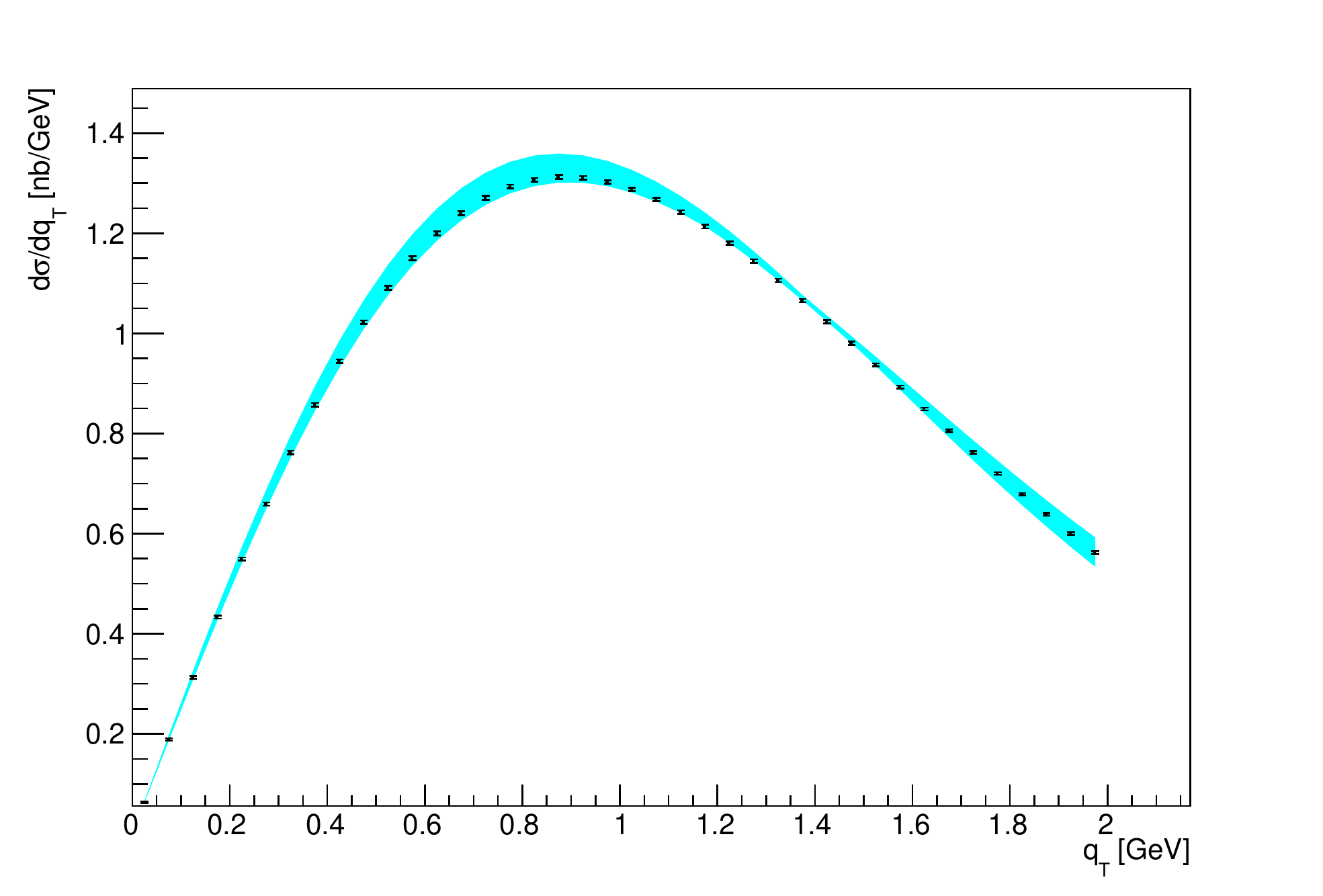}
\caption{N$^3$LL theory curve from Ref.~\cite{Gutierrez-Reyes:2019vbx} compared to statistical projections from Belle~II using 10~fb$^{-1}$ of simulated data. The analysis uses a jet radius of $R=1.0$ and a requirement that the jet energy is larger than 3.75 GeV, corresponding to $z=0.713$. Belle~II data will allow a multidimensional measurement, e.g., in $z$, jet radius and $q_T$, of this quantity which will be systematically limited.\label{fig:jetQtMC} }
\end{figure}

An important aspect of jet physics are studies of transverse momentum dependent (TMD) processes. For example, at the EIC it will be possible to study the structure of nucleons and nuclei in terms of TMD PDFs using jet correlations~\cite{Gutierrez-Reyes:2018qez,Liu:2018trl,Gutierrez-Reyes:2019vbx,Kang:2021ffh}. An important ingredient are TMD fragmentation and jet functions, which can be measured in great detail at Belle~II. This is relevant for both spin polarized and unpolarized scattering processes. For example, in Ref.~\cite{Gutierrez-Reyes:2018qez} it was proposed to measure di-jet correlations in $e^+e^-$ collisions. The factorization structure involves universal jet functions, which are also relevant for the corresponding factorization theorems in electron-proton collisions at the future EIC. While the jet and soft functions are perturbative objects, they are nevertheless sensitive to hadronization corrections. By performing the corresponding measurements at Belle~II, the non-perturbative component can be constrained and the data can be incorporated in global fits of TMD PDFs and fragmentation/jet functions. An illustration of the di-jet correlation observable and numerical predictions up to N$^3$LL accuracy for $e^+e^-$ collisions at Belle~II are shown in Fig.~\ref{fig:dijet}. Figure~\ref{fig:jetQtMC} shows statistical projections from Belle~II data for this observable. Belle~II statistics will allow a fine multidimensional binning.
Moreover, in Ref.~\cite{Liu:2021ewb}, time-reversal odd (T-odd) aspects were discussed in the context of jet physics. Cross sections sensitive to T-odd effects involving jets are non-vanishing due to non-perturbative effects. The T-odd jet function is related to the Collins fragmentation function~\cite{Collins:1992kk}. It can be accessed through anisotropies in di-jet correlations in $e^+e^-$ collisions at Belle~II. Similar cross sections can be measured at the future EIC. Precision studies in $e^+e^-$ collisions at Belle~II will undoubtedly play an important role in improving our understanding of these recently proposed observables.

\begin{figure}[t]
\centering
\includegraphics[width=0.99\textwidth]{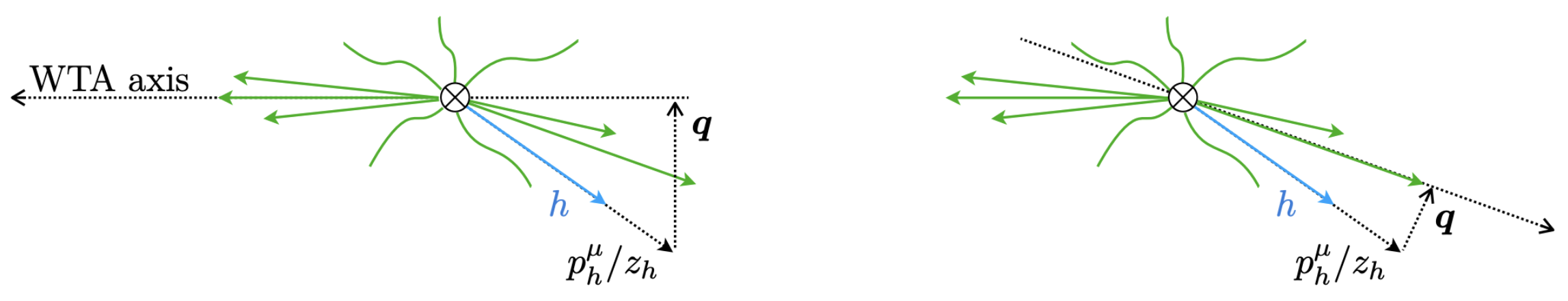}
\caption{Processes to measure TMD fragmentation functions in $e^+e^-$ collisions. Left: The Winner-Take-All (WTA) axis is identified in one hemisphere and the hadron transverse momentum spectrum is measured in the opposite hemisphere. Right: Jets are identified with the standard or WTA axis and the hadron transverse momentum spectrum is measured with respect to either of the two axes. Figure taken from Ref.~\cite{Makris:2020ltr}.~\label{fig:TMDFF}}
\end{figure}

Next, we discuss three different methods to extract TMD fragmentation functions in $e^+e^-$ collisions. First, one can measure the transverse momentum of hadrons inside jets with respect to the standard jet axis~\cite{Procura:2009vm,Kaufmann:2015hma,Bain:2016rrv,Kang:2017glf}. This observable has been measured in in proton-proton collisions at the LHC and RHIC. Corresponding measurements at Belle~II would provide constraints in a different kinematic regime and allow for a separation of quark and gluon jets. Second, instead of using the standard jet axis, jets can be clustered with the Winner-Take-All (WTA) axis. The WTA axis is recoil-free and the corresponding transverse momentum spectrum of hadrons with respect to the WTA axis is insensitive to soft radiation. For example, it eliminates non-global logarithms. As a result, the standard TMD fragmentation functions do not appear in the corresponding factorization theorem. Instead, the single logarithmic series is resummed by a modified DGLAP evolution equation~\cite{Neill:2016vbi,Neill:2018wtk}. Third, in Ref.~\cite{Makris:2020ltr} it was proposed to identify the WTA axis in one hemisphere in $e^+e^-$ collisions and to measure the transverse momentum of hadrons with respect to this axis in the opposite hemisphere. This approach has the advantage that it does not rely on an additional measurement like thrust to enforce the di-jet limit and, in addition, there are no non-global logarithms that limit the perturbative accuracy that can be achieved. As an example, we give the factorization theorem for this cross section differential in the hadron momentum fraction $z_h$ and its transverse momentum $\boldsymbol{q}$: 
\begin{equation}
    \frac{\mathrm{d} \sigma}{\mathrm{d} z_{h} \mathrm{d} \boldsymbol{q}}=\sigma_{0} \int_{-\infty}^{\infty} \frac{\mathrm{d} \boldsymbol{b}}{(2 \pi)^{2}} e^{\mathrm{i} \boldsymbol{b} \cdot \boldsymbol{q}} H(Q, \mu)\, J\left(\boldsymbol{b}, \mu, \frac{\nu}{Q}\right) S(\boldsymbol{b}, \mu, \nu) \sum_{j} D_{j \rightarrow h}\left(z_{h}, \boldsymbol{b}, \mu, \frac{\nu}{Q}\right) \,.
\end{equation}
Here, $H$, $J$ and $S$ are hard, jet and soft functions in Fourier transform space, respectively, and D represents the standard TMD fragmentation function that can be constrained by comparing to future measurements at Belle~II. See Ref.~\cite{Makris:2020ltr} for more details. The three different processes discussed here to extract transverse momentum dependent fragmentation functions are illustrated in Fig.~\ref{fig:TMDFF}.

Instead of measuring the transverse momentum spectrum of hadrons with respect to different jet axes, Ref.~\cite{Cal:2019gxa} introduced the angle between different jet axes as a novel observable to constrain TMD dynamics. Different than hadron spectra, the angle (or transverse momentum) between jet axes is an Infrared-Collinear (IRC) safe observable and as such independent of collinear fragmentation functions. However, it was shown that this observable is nevertheless directly sensitive to TMD dynamics and in particular the non-perturbative component of the rapidity anomalous dimension. For example, lattice QCD can constrain the rapidity anomalous dimension~\cite{Ebert:2018gzl}. Recently, the ALICE Collaboration at the LHC presented first results for the angles between the standard, WTA and groomed jet axes~\cite{Mulligan:2021gyl}. High-precision measurements at Belle~II would constrain the rapidity anomalous dimension and within a global analysis it would allow for the separation of quark and gluon contributions.

There is a range of other jet-related observables that constrain different QCD dynamics. Corresponding high-precision measurements at Belle~II would greatly improve our understanding of the limits of perturbative QCD and constrain relevant nonperturbative quantities. Examples include energy-energy correlators~\cite{Verbytskyi:2019ikn,Dixon:2019uzg}, the momentum sharing fraction $z_g$~\cite{Larkoski:2015lea,Cal:2021fla}, the jet charge~\cite{Waalewijn:2012sv,Kang:2020fka}, flavor correlations~\cite{Chien:2021yol}, jet pull~\cite{Larkoski:2019urm,Larkoski:2019fsm}, and jet angularities~\cite{Lee:2006nr,Ellis:2010rwa,Aschenauer:2019uex,Lee:2019lge,Caletti:2021oor}.

\subsection{Correlation studies}
\label{sec:correlations}
As already anticipated several times in this document, the next step in our understanding of hadronization dynamics requires a move from more inclusive measurements to correlation measurements. At the LHC such a program is already underway, mainly focusing on jet physics. A similar program in the clean $e^+e^-$ environment focusing on particle-based observables and at at a lower center of mass energy, providing a lever arm in energy, will enable tests of our understanding of QCD dynamics, that do not have an alternative. 
This section will describe two measurements in detail which are designed to understand hadronization dynamics by correlations. In subsection~\ref{sec:leadcorr} flavor correlations of leading hadrons, for which currently no factorization scheme is available and in subsection~\ref{sec:dihadron} the partial wave decomposition of di-hadron correlations, which aims at a complete description of two-hadron correlations in the final state. For di-hadron fragmentation functions, a factorization theorem is available and the different partial waves are associated with different interferences on the QCD amplitude level. This interpretability is another attractive aspect of these objects.

\subsection{Probing hadronization with flavor correlations of leading particle in jet.}
\label{sec:leadcorr}
The evolution of energetic partons, namely quark and gluons, into final state particles involves both perturbative and nonperturbative phenomena. Many perturbative aspects of jets can be studied at high energy colliders like LHC, RHIC, and LEP. Jets at EIC are more towards nonperturbative regions while Belle~II is an ideal place where one can study nonperturbative phenomena in full fledged. The work~\cite{Chien:2021yol} emphasizes an unique way of accessing the nonperturbative region using the charge and momentum correlations of the leading particles of a jet. The observable charge-correlator, $r_c$, is expected to be sensitive to the dynamics of hadronization, whether it is of string-like or of random correlations from very naive picture of quark and anti-quark sitting at the end of strings. The strength of $r_c$ combining specific flavor correlation can be explained with simple string connections to form final state particles. Particle identification at high momentum is essential to make such studies. Belle~II has an extensive capability of flavor tagging to measure the correlations.
Belle~II can make good measurements with formation time~\cite{Dokshitzer:1991wu} $t_{form}=z(1-z)p/k_\perp^2$, relative transverse momentum of the two leading particles and with other kinematic variables. Other than the proposed experiments at EIC, no other current experiment can measure the flavour dependency, and therefore, the measurements would be very valuable to formulate a theory of fragmentation if we are very optimistic. Here is the charge correlation ratio, $r_c$, from the differential cross sections ${\rm d}\sigma_{H_1H_2}/{\rm d}X$ to quantify flavor and kinematic dependence of hadronization in the production of $H_1=h_1$ and $H_2=h_2$ or $\overline{h_2}$ (the antiparticle of $h_2$),
\begin{equation}
	r_c (X)= \frac{{\rm d}\sigma_{h_1h_2}/{\rm d}X-{\rm d}\sigma_{h_1\overline{h_2}}/{\rm d}X}{{\rm d}\sigma_{h_1h_2}/{\rm d}X+{\rm d}\sigma_{h_1\overline{h_2}}/{\rm d}X}\;.
	\label{eq:r_c-def}
\end{equation}
In the definition, Eq.\ (\ref{eq:r_c-def}), $H_1$ and $H_2$ can in principle be arbitrary hadron species, including charged and neutral hadrons. 

Here we will demonstrate with PYTHIA-8 event generator to extract $r_c$ and compare Belle~II with LEP and EIC. Belle~II simulations for $\epem$ collisions at 10.52 GeV center of mass energy (a 7 GeV electron colliding with a 4 GeV positron) are done where jets are produced from u, d, s, c and their anti quarks. Though the multiplicity and the energy of the jets at Belle~II is far from the region where one can think of perturbative calculations, we used jet reconstruction to get the constituents in jets and with fewer particles in jets it might be advantageous to understand the dynamics of fragmentation making a clear physical picture.  A spherically invariant fastjet package~\cite{Cacciari:2011ma} used to reconstruct jets where the distance calculated in terms of the opening angle between the two pseudo-jets. Without any detailed detector acceptance and reconstruction limitations, the analysis is done at the particle level to gauge the correlations embedded in the fragmentation model of PYTHIA. Particles selected with  transverse momentum $p_{T}>0.2$ GeV with $-3.0 <\eta < 3.0$. The jet radius is taken as $R=0.8$. At $\sqrt{s}$ = 91.0 GeV  LEP collide a positron and an electron. For comparison with DIS jets PYTHIA-8 simulations with 18 GeV election with 275 GeV proton is made. The standard anti-kt algorithm with $R=1.0$ is used for EIC.
\begin{figure}[h!]
\centering
\includegraphics[width=0.99\textwidth]{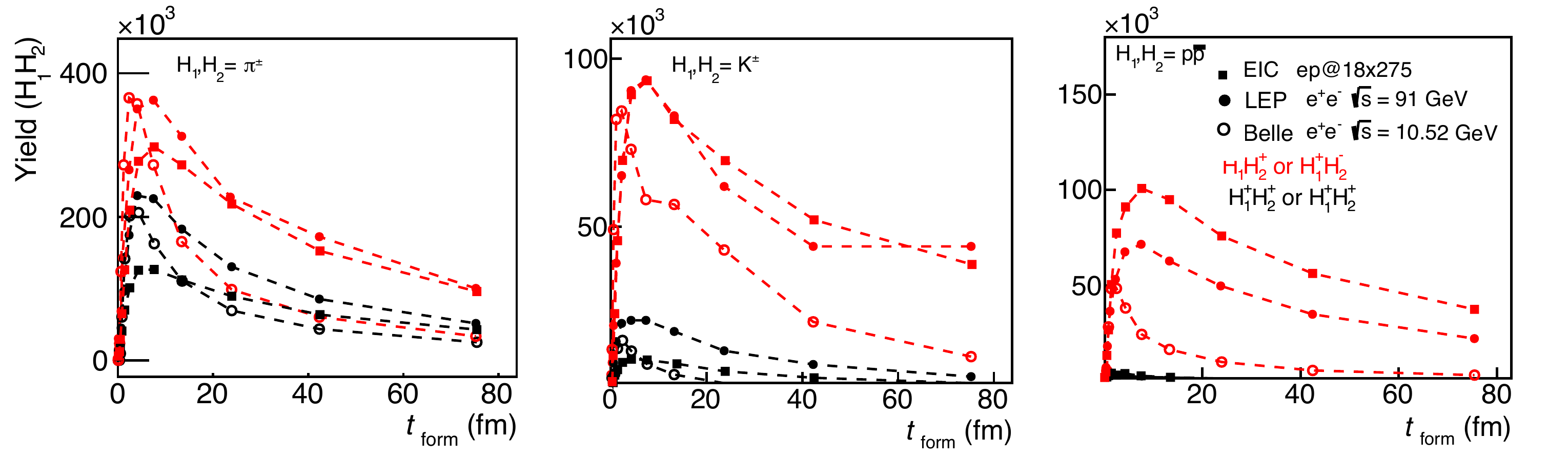}
\caption{Yield of $H_{1}H_{2}$ pairs for charged pions, charged kaons, protons and antiprotons with formation time. Leading and Next to leading pairs of same charges are in BLACK and opposite charges are in RED. Yields of opposite charge pairs are large and it is dependent on particle species of pairs chosen. Formation time for LEP and EIC are comparable, while in case of Belle the peak of the formation time appears earlier.}
\label{fig:ft}
\end{figure}
\begin{figure}[h!]
\centering
\includegraphics[width=0.92\textwidth]{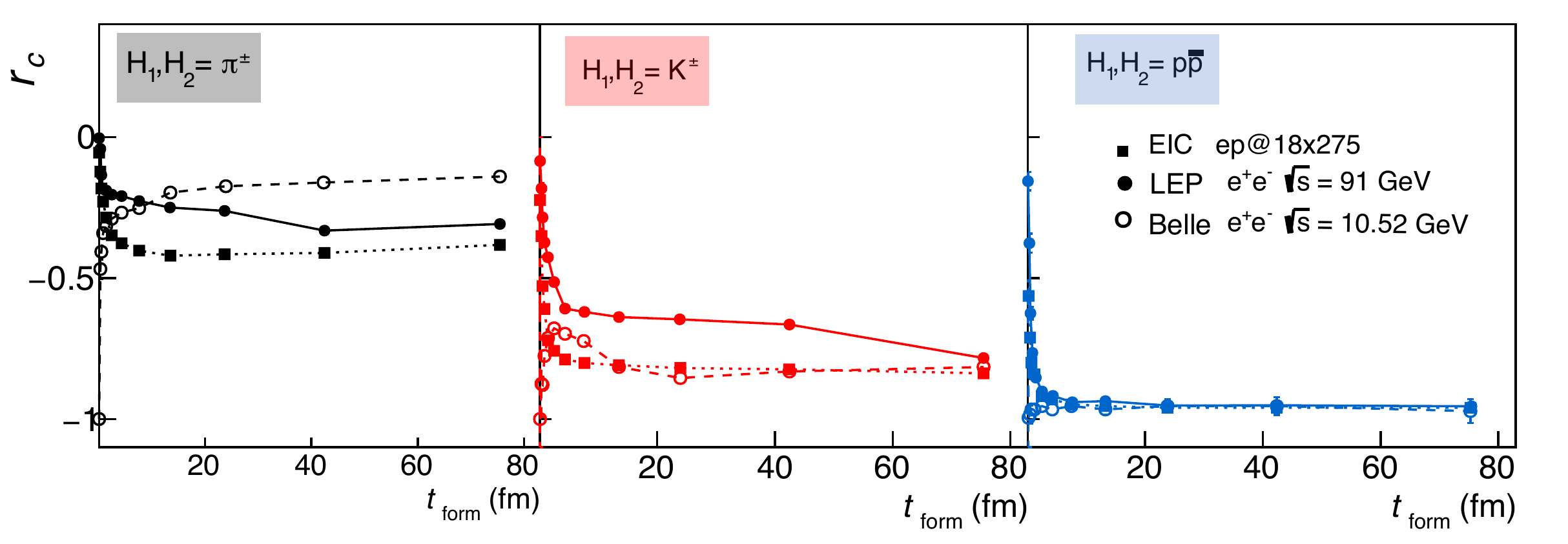}
\caption{{$\it {r_c}$} with formation time ($t_{form}$) for different flavor species. There are strong flavor dependent correlations and EIC and LEP correlations at small formation time seem to be different from Belle. Perturbative process is dominant at small formation time at LEP and EIC while at Belle the mechanism of fragmentation might be fully nonperturbative in nature.}

\label{fig:rc_f}
\end{figure}

\begin{figure}[h!]
\centering
\includegraphics[width=0.92\textwidth]{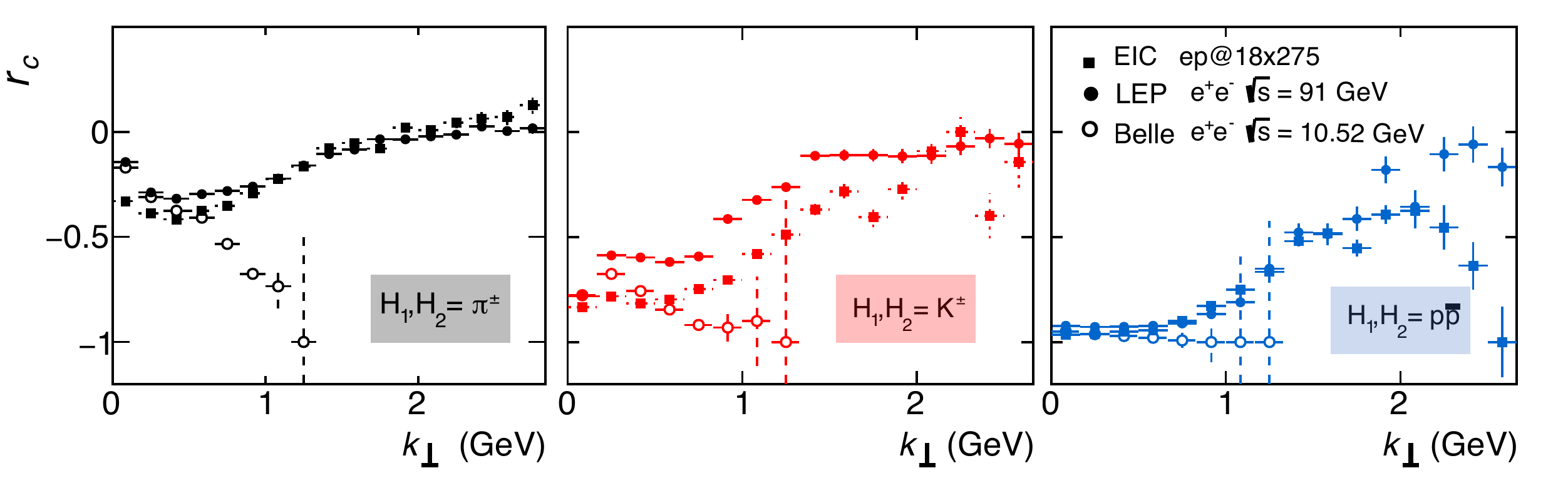}
\caption{{$\it {r_c}$} with $k_{\perp}$ (transverse momentum  of leading or next-to-leading with respect to their resultant). Belle $k_{\perp}$ don't reach at high values. {$\it {r_c}$} for Belle shoes very distinct nature of falling  with $k_{\perp}$. This might be be an indication of gluon radiation which is not a dominant part of evolution of jet.}
\label{fig:rc_k}
\end{figure}
Formation time in the order of a few fm is the region where perturbative processes dominate. Then the transition region lies in a few tens of fm and finally the large formation time indicates the region dominated mostly by nonperturbative fragmentation. The Fig.~\ref{fig:ft} shows the formation time distributions for Belle has early and rapidly falling peak for $H_{1}H_{2} \pi^{\pm}$ or $K^{\pm}$ or $p\bar{p}$. The yield of opposite charge pairs are smaller compared to the same charge pairs suggesting the charge ordering in momentum embedded in the event generators. Since formation time can separate regions with the nature of perturbation applied, a comparative measurement with Belle with high energy jets would be suggestive to understand the fragmentation dynamics. Figure~\ref{fig:rc_f} shows the charge correlation $r_c$ for different flavor combinations where the strength of correlations have strong flavor dependency. $r_c$ values for Belle at small formation time dips down while for EIC and LEP it is rising and large angle perturbative radiations might be making the difference which might be relatively frequent at LEP and EIC compared to that of Belle. Measurement with other kinematic variables would also be very interesting at Belle to understand the dynamics of fragmentation. One example is with $k_{\perp}$ as shown in Fig.~\ref{fig:rc_k}. Large $k_{\perp}$ can come for early gluon radiation and the decorrelation in $r_c$ can be seen for LEP and EIC while for Belle large $k_{\perp}$  region absent. The difference of the $r_c$ for Belle and LEP or EIC is significant and in particular the falling trend in Belle with increasing $k_{\perp}$ which is opposite for LEP and EIC.\\

The specific flavor tagging explained in ~\cite{Chien:2021yol} is sensitive to string picture of fragmentation where the leading particle is $\pi^{+}$ or $\pi^{-}$ and in combination with next-leading particle with $K^{\pm}$. Belle would potentially be able to make such favor tagging measurements with u and d quark tagging and it can reveal deep insight into the string fragmentation picture. An extensive scale of correlations can be studied for various configurations and even we can extend the correlations between the two jets originates from quark and antiquark origin in an event.


\section{Determination of $\alpha_S$}
\label{sec:alphas}

The strong coupling $\alphas$ quantifies the strength of the strong interaction among quark and gluons at any given energy, and is one of the fundamental parameters of the SM. Its current value at the reference Z pole mass amounts to $\alphasmZ=0.1179 \pm 0.0009$~\cite{Zyla:2020zbs}, with a $\delta\alphas/\alphas \approx 0.8\%$ uncertainty that is orders of magnitude larger than that of the other three interaction (QED, weak, and gravitational) couplings. Improving our knowledge of $\alphas$ is crucial, in particular, to reduce the theoretical ``parametric'' uncertainties in the calculations of all pQCD observables (cross sections, decay rates, masses, etc.\ for multiple particles) that depend on an expansion in powers of $\alphas$. The current $\alphasmZ$ uncertainties propagate and percolate through multiple high-precision Higgs, electroweak, flavor, or top physics studies at current and future colliders, of relevance for direct and indirect searches for physics beyond the SM~\cite{Proceedings:2019pra,Proceedings:2019vxr}. 

A detailed study of the current status and future prospects of $\alphas$ determinations has appeared recently~\cite{dEnterria:2022hzv}. Among the seven currently used categories of measurements used to extract the world-average QCD coupling constant in the PDG~\cite{Zyla:2020zbs}, there are three that would clearly benefit from the state-of-the-art measurements of various observables accessible at Belle~II, by exploiting its very large data samples collected as well as its reduced systematic uncertainty compared to similar studies at previous $\epem$ facilities. Among the needed $\alphas$-sensitive measurements at Belle-II, one can emphasize:
\begin{itemize}
    \item $\tau$ lepton spectral functions;
    \item $\epem$ R-ratio over $\sqrt{s} \approx 2$--10~GeV;
    \item $\epem$ event shapes ($C$-parameter, thrust, energy-energy correlators,...);
    \item parton-to-hadron fragmentation functions;
\end{itemize}
We succinctly cover these four topics below.

\subsection{\texorpdfstring{Hadronic $\tau$}{tau} decays}

The inclusive distribution of the final hadrons (spectral functions) measured in $\tau$ decays at LEP~\cite{OPAL:1998rrm,ALEPH:1998rgl,ALEPH:2005qgp} provides a precise determination of the strong coupling at the lowest energies accessible, through comparisons of data to pQCD calculations~\cite{Pich:2016bdg,Boito:2016oam,Ayala:2021mwc} known to $\mathcal{O}(\alphas^4)$ accuracy~\cite{Baikov:2008jh}. The present $\alphasmZ = 0.1178 \pm 0.0019$ value of the $\tau$ category has a $\pm 1.6\%$ uncertainty as derived from the averaging of four different extractions based on two different theoretical perturbative approaches: contour-improved (CIPT) or fixed-order (FOPT) perturbation theory. New measurements of the $\tau$ spectral functions with lower (statistical and systematical) uncertainties will allow for a reduction in (i) the experimental part of the $\alphas(m_\tau^2)$ propagated uncertainty, as well as in (ii) the nonperturbative power corrections of the calculations, which are nonnegligible given the relatively low energy scale of the tau mass. On the nonperturbative side, increased precision in the data would allow more accurate subtractions of the (model-dependent) quark-hadron duality-violation contributions. In particular, more precise 2-pion and 4-pion exclusive-mode $\tau$ experimental data would help to better disentangle perturbative and nonperturbative contributions.

While the FCC-ee would ultimately produce enormous $\tau$-lepton data samples~\cite{FCC:2018evy}, the best near-future prospect~\cite{dEnterria:2022hzv} is provided by Belle-II, which has access to many more $\tau$-leptons than were produced at LEP. No fully-inclusive spectral function needs to be obtained from Belle-II measurements, but more precise 2-pion and 4-pion exclusive-mode $\tau$ data would already help to produce a new tau-lepton vector spectral function with smaller uncertainties than available today~\cite{Boito:2020xli}. Details will need to be appropriately considered. For instance, while the already existing Belle unit-normalized 2-pion distribution is more precise than that of ALEPH or OPAL, the $\tau\to\pi^-\pi^0\nu_\tau$ channel has been measured less well by Belle, with the HFLAV~\cite{HFLAV:2019otj} value still dominated by ALEPH. An improved, branching-fraction-normalized 2-pion distribution will thus require combining input from different experiments. The situation is, presumably, similar for the two 4-pion modes.

\subsection{\texorpdfstring{$R$-ratio over $\sqrt{s} \approx 2$--10~GeV}{R-ratio over sqrt(s) = 2--10 GeV}}

The cross section for $\epem$ annihilation into hadrons, both at resonance poles ($\tau$, W, Z) as well as in the continuum, provides very clean ways for measuring the strong coupling constant. The ``timelike'' experimentally measurable observable $R$-ratio of $\epem$ annihilation into hadrons at a given center-of-mass energy $\sqrt{s}$, $R(s) = \sigma(e^{+}e^{-} \!\to \text{hadrons}; s)/\sigma(e^{+}e^{-} \!\to \mu^{+}\mu^{-}; s)$, can be connected with ``spacelike'' theoretically computable quantities such as the hadronic vacuum polarization function and the Adler function, which can be written as a perturbative expansion of $\alphas$ powers~\cite{Nesterenko:2017wpb}. In principle, a wide range of energies exists to measure $R(s)$ ---ranging from the region below the charm threshold up to the highest energies available at future electron-positron colliders--- which can be exploited to extract the strong coupling. At low energies, the CLEO data was used to extract $\alphas$ albeit with relatively poor precision~\cite{Kuhn:2007tc}. More recently, all $R$-ratio data below $\sqrt{s} = 2$~GeV was analyzed to complement the QCD coupling extraction based on the $\tau$-lepton results~\cite{Boito:2018yvl}. More precise measurements of $R(s)$ over $\sqrt{s}\approx 1$--10~GeV at Belle-II through the radiative return technique explained in Section~\ref{sec:g-2}, can certainly be similarly exploited to provide new independent $\alphas$ determinations.


\subsection{\texorpdfstring{$\epem$}{e+e-} event shapes}
\label{sec:eecorr}

Event-shape variables in $\epem$ collisions ---such as the thrust, $C$-parameter, and energy-energy correlators--- have been measured with high precision by the LEP experiments. These observables have played and continue to play an important role in the determination of the value of the strong coupling through pQCD analysis at N(N)LO+N(N)LL accuracy~\cite{Dissertori:2009ik, OPAL:2011aa,Gehrmann:2012sc,Kardos:2020igb}.
The present precision of the $\alphasmZ = 0.1171 \pm 0.0031$ determination from the $\epem$ event-shapes and jet-rates category of the PDG world-average is of $\pm2.6\%$. Such a relatively large uncertainty is driven by the large span among $\alphasmZ$ values derived from measurements where nonperturbative corrections have been obtained with Monte Carlo parton showers (PS) or analytically. Three parallel axis have been suggested to reduce this current uncertainty: (i) extend the latest analytical studies of nonperturbative power corrections~\cite{Caola:2022vea}, and improve the PS algorithms to reach NNLL accuracy~\cite{Dasgupta:2020fwr} and match them to NNLO predictions, (ii) employ soft-drop grooming techniques to the $\epem$ data to evaluate their impact in the reduction of nonperturbative effects~\cite{Marzani:2019evv}, and (iii) incorporate new $\epem$ event-shapes measurements with reduced statistical and systematic uncertainties compared to the existing LEP studies. In this last front, the modern detector systems of Belle~II and its huge hadronic data samples can certainty provide novel useful event-shapes measurements from which precise $\alphas$ extractions can be performed.

As already mentioned in Sec.~\ref{sec:jets}, energy-energy correlators are a very interesting event shape variable to study, due to their connection to TMD functions and their precise knowledge from perturbative QCD calculations at N$^3$LL~\cite{Moult:2018jzp}. 
For details we refer to the dedicated whitepaper in Ref.~\cite{Neill:2022lqx}. We emphasize here also that Belle~II will operate in a kinematic regime not yet explored for these correlators, which can be in addition exploited to carry out $\alphas$ extractions (as done at NNLO+NNLL accuracy with LEP data in~\cite{Kardos:2020igb}). The lower energy of Belle~II compared to, e.g., LEP will enlarge the current uncertainty somewhat, in the order of 5--10\% with current calculations compared to a few percent at LEP, but 
the Belle~II energy point and the experimental precision of the measurements expected, will make them very interesting analyses.


\subsection{Parton-to-hadron fragmentation functions}

The QCD coupling has been extracted ---up to approximate next-next-to-leading-order (NNLO$^\star$) fixed-order including next-to-next-to-leading-log (NNLL) corrections (resummed in the so-called next-to-MLLA approach~\cite{Dokshitzer:1991wu})--- through the study of the energy evolution of the moments of the parton-to-hadron inclusive FFs at low hadron Feynman momentum fraction $z$~\cite{Perez-Ramos:2013eba,dEnterria:2014xmb,Perez-Ramos:2019zxf}. 
The QCD coupling obtained from the combined fit of FF multiplicity and peak position measured in $\epem$ over the $\sqrt{s}\approx$~2--200~GeV range, is $\alphasmZ = 0.1205 \pm 0.0010\,(\mathrm{exp})\,\pm 0.0022\,(\mathrm{theo})$, where the first uncertainty includes all experimentally-related sources and the second one is the theoretical scale uncertainty. Although the dominant source of imprecision is of theoretical nature (and reducing it requires to match the MLLA and $\MSbar$ anomalous dimension and to reach full-NNLO pQCD accuracy), the propagated experimental uncertainties of $\sim$0.8\% can be further reduced by exploiting the larger and more precise data samples that can be collected at Belle-II over $\sqrt{s}\approx$~2--10~GeV.

Apart from the study of the soft FFs, the QCD coupling can be also determined from fits of the scaling violations of the inclusive hard (high-$z$) FFs~\cite{Proceedings:2015eho}. Strong coupling extractions from the hard FFs exist at NLO accuracy~\cite{Albino:2005me}, which was the state-of-the-art up to a few years ago. The recent availability of global FF analyses at NNLO accuracy~\cite{Anderle:2015lqa,Soleymaninia:2018uiv,Borsa:2022vvp}, combined with improved Belle-II measurements, can lead to new extractions of $\alphasmZ$ with few percent uncertainties.

\section{Dynamical mass generation studies}
\label{sec:jet_mass}

\newcommand{\psibar}{{\overline{\psi}}}
\newcommand{\id}{{\mathbb{I}}}
\newcommand{\vect}[1]{\vec{#1}} 

\begin{figure}[tb]
	\centering
	\includegraphics[width=0.36\linewidth]{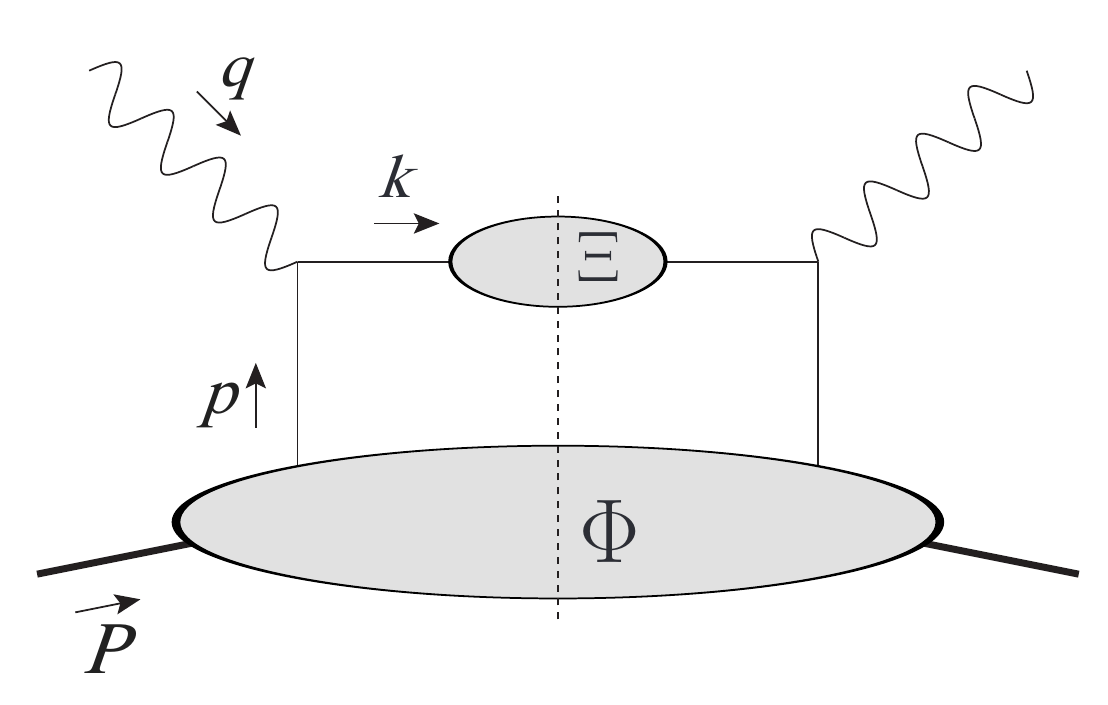}
	\quad
	\includegraphics[width=0.47\linewidth]{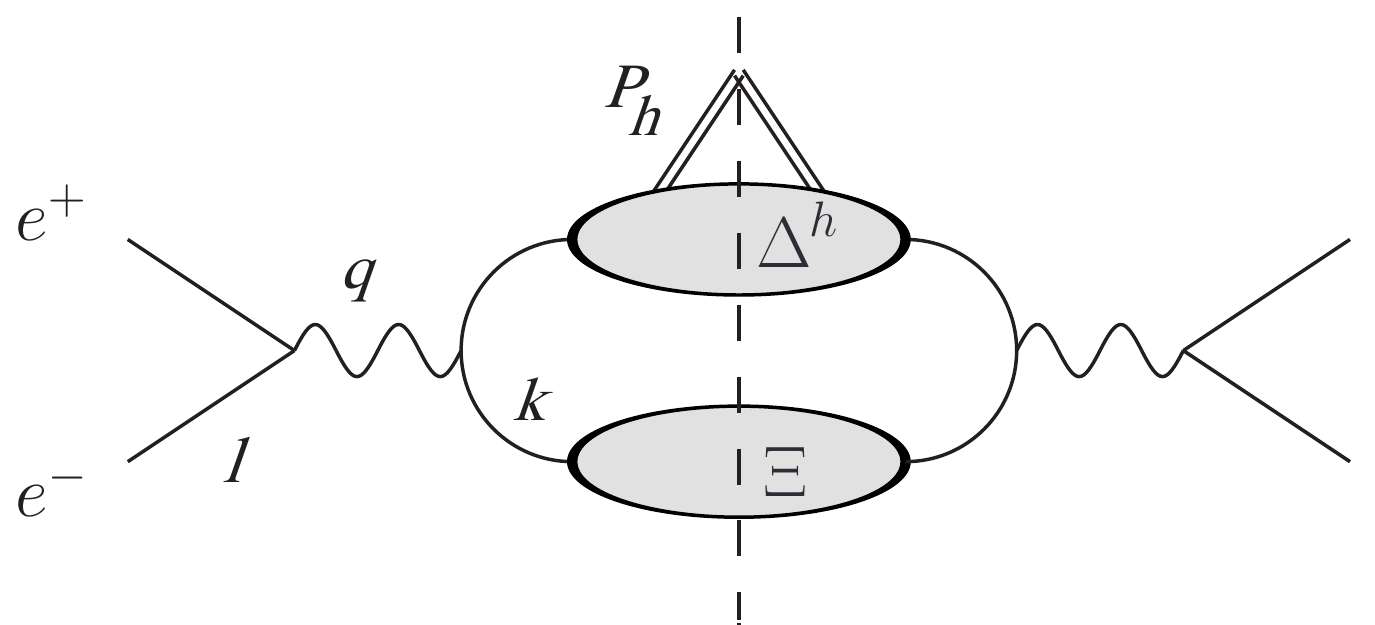}
	\caption{Inclusive deep-inelastic scattering (left) and $\Lambda$ production from semi-inclusive annihilation (right) diagrams with an inclusive jet correlator $\Xi$ replacing unobserved perturbative quarks in the final state. 
	$\Phi$ and $\Delta^h$ are the correlation functions that encode information on hadron structure and hadronization, respectively. 
	In this document, the detected hadron $h$ is a self-polarizing $\Lambda$ baryon.}
	\label{fig:jetmass_processes}
\end{figure}

The proposed upgrade of SuperKEKB to include polarized beam with a polarization of up to 70\%~\cite{polarizedSuperKEKBWhitepaper}, opens up additional possibilities in the study of QCD. Investigating dynamical mass generation is one exciting possibility, which will be further explored in this section.
Due to color confinement, the quarks created in hard collisions cannot appear as on-shell particles in the final state, but rather decay into a jet of hadrons whose mass is dynamically generated. 
The details of the  quark-to-hadron transition are still unknown. As proposed in Ref.~\cite{Accardi:2019luo,Accardi:2020iqn}, dynamical mass generation can be studied even without observing the produced hadrons, by analyzing the chiral-odd component of the (color averaged) \textit{gauge-invariant} quark correlator
$
    \Xi_{ij}(k) = {N_c^{-1}}\text{Tr}_c\, \text{Disc} \int \frac{d^4 \xi}{(2\pi)^4} e^{i k \cdot \xi} \,
    \langle\Omega| \psi_i(\xi) \psibar_j(0) W(0,\xi;n_+) |\Omega\rangle
$,
where $|\Omega\rangle$ is the nonperturbative QCD vacuum, $\psi_i$ is the quark field, $W$ a Wilson line. This correlator describes the nonperturbative propagation and hadronization of a quark \cite{Accardi:2019luo,Accardi:2020iqn}. When integrated over the subdominant quark momentum component $k^+$, the resulting inclusive jet correlator 
\begin{align}
    J_{ij}(k^-,\vect{k}_T)
    & \equiv \frac{1}{2} \int dk^+\, \Xi_{ij}(k) 
    \nonumber \\
    & = \frac{\theta(k^-)}{4(2\pi)^3\, k^-} \, 
    \bigg\{ k^-\, \gamma^+ + \slashed{k}_T + M_j \id + \frac{K_j^2 + \vect{k}_T^2}{2k^-} \gamma^- \bigg\}
\label{eq:ijet_correlator}
\end{align}  
generalizes the perturbative quark propagator contributing to particle production in lepton-nucleus DIS scattering at large Bjorken $x$ values~\cite{Accardi:2008ne,Accardi:2017pmi}, as well as in the semi-inclusive annihilation (SIA) of electrons and positrons, see Fig.~\ref{fig:jetmass_processes}. Note that in Eq.~\eqref{eq:ijet_correlator} we assume that $k^- \sim Q \gg |\vect{k_T}|\gg k^+$, with the hard scale $Q$ of the process providing one with a ``twist'' expansion of the jet correlator. 
The twist-4 $K_j^2$ coefficient quantifies the invariant mass of the unobserved quark hadronization products. 
The twist-3 $M_j = m_q + m_q^{dyn}$ coefficient -- also called ``jet mass'' -- quantifies the mass acquired by the quark during its hadronization, and is composed of the explicit current quark mass term $m_q$ and a dynamically generated term $m_q^{dyn}$.
Crucially, the jet mass can be calculated as the integral of the quark's chiral-odd spectral function $\rho_1$: 
\begin{align}
\label{e:Mj_def}
    M_j = \int_0^\infty d\mu^2 \sqrt{\mu^2}\,\rho_1(\mu^2) 
\end{align}
where the sum runs over the flavor and spin of the hadrons produced with mass $M_h$. 

In inclusive DIS, $M_j$ couples to the chiral-odd leading twist transversity distribution $h_1(x)$ of the proton target and contributes to the LT double spin asymmetry, in particular, to the $g_2$ structure function~\cite{Accardi:2017pmi}. 
Comparing this to the JAM15 polarized DIS global fit indicates $M_j \approx 0.1$ GeV at perturbative scales. 
Evolving this back to an initial $Q_0=0.6$ GeV scale one finds $M_{j0} \approx 0.5$ GeV.

When producing a self-polarizing $\Lambda$ hadron in the SIA of polarized electrons and positrons, the dynamical component of the jet mass $M_J$ couples to the chiral-odd twist-3 transversity fragmentation function $H_1$ :
\begin{align}
\label{e:dsigma_L}
    \frac{d\sigma^L}{d\Omega\, dz} 
    & = \frac{3 \alpha^2}{Q^2}\, \lambda_e\, \sum_a e_a^2 \, 
    \bigg\{ 
        \frac{1}{2}\, \lambda\, C(y)\, G_1^{a \to \Lambda}(z,Q)  
        \nonumber \\
    & + 2 D(y)\, |\bm{S}_T|\, \cos(\phi)\, \frac{M_\Lambda}{Q} 
        \bigg( \frac{1}{z}\, G_T^{a \to \Lambda}(z,Q)
        + \frac{m_a^\text{dyn}}{M_\Lambda}\, H_1^{a \to \Lambda}(z,Q) \bigg)
    \bigg\} \ . 
\end{align}
The jet mass can be accessed measuring the longitudinal $A_L$ electron spin asymmetry
\begin{align}
\label{e:asymmetry}
    A_L & = \frac{d\sigma^\rightarrow - d\sigma^\leftarrow}{d\sigma^\rightarrow + d\sigma^\leftarrow} \ ,
\end{align}
where ``L'' refers to the longitudinal lepton polarization. 
The $\Lambda$'s longitudinal and transverse spin contributions can be separated studying the $y = P_\Lambda \cdot l / P_\Lambda \cdot q$ dependence of the asymmetry, where $l$, $q$, and $P_\Lambda$ are the four-momenta of the incoming electron, the exchanged photon, and the $\Lambda$ baryon respectively~\cite{Boer:1997mf,Boer:2008fr}. 
The coefficients $A(y)$, $C(y)$, $D(y)$ can be found in Ref.~\cite{Boer:1997mf,Boer:2008fr}.

Lacking, at this time, experimental information on the polarized fragmentation functions entering Eq.~\eqref{e:dsigma_L}, we use positivity bounds, QCD equations of motion and the WW approximation-\cite{Metz:2016swz} to relate these functions to the known unpolarized $D_1$ fragmentation function of the $\Lambda$ hadron and obtain an order of magnitude estimate for the asymmetry\footnote{We note, however, that a measurement of the $G_1^\Lambda$, $G_T^\Lambda$ and $H_1^\Lambda$ FFs can be obtained by detecting another hadron or a di-hadron pair in the hemisphere opposite to the $\Lambda$.}. 
The positivity bounds for $G_1$, $H_1$ are
\begin{align}
\label{e:G1_H1_bounds}
    |G_1(z)| \leq D_1(z) \, ,
    \quad 
    |H_1(z)| \leq D_1(z) \, ,
\end{align}
and the EOM for the twist-three $G_T$ FF read~\cite{Mulders:1995dh}:
\begin{align}
\label{e:GT_EOM}
    & G_T(z) = z\, G_{1T}^{(1)}(z) + \tilde{G}_T(z) + \frac{m}{M_\Lambda}\, z\, H_1(z) \, , 
\end{align}
where $M_\Lambda$ is the mass of the $\Lambda$ hadron, $m$ is the current mass of the fragmenting quark, and $G_{1T}^{(1)} = \int d^2 \bm{k}_T\, \frac{\bm{k}_T^2}{2 M_h^2}\, G_{1T}(z,\bm{k}_T^2)$. The WW approximation amounts to neglecting the dynamical twist-three function $\tilde{G}_T(z)$. We also neglect $G_{1T}^{(1)}(z)$, since we do not have any information on the size of this term.  

\begin{figure}[tbh]
	\centering
	\includegraphics[width=0.49\linewidth]{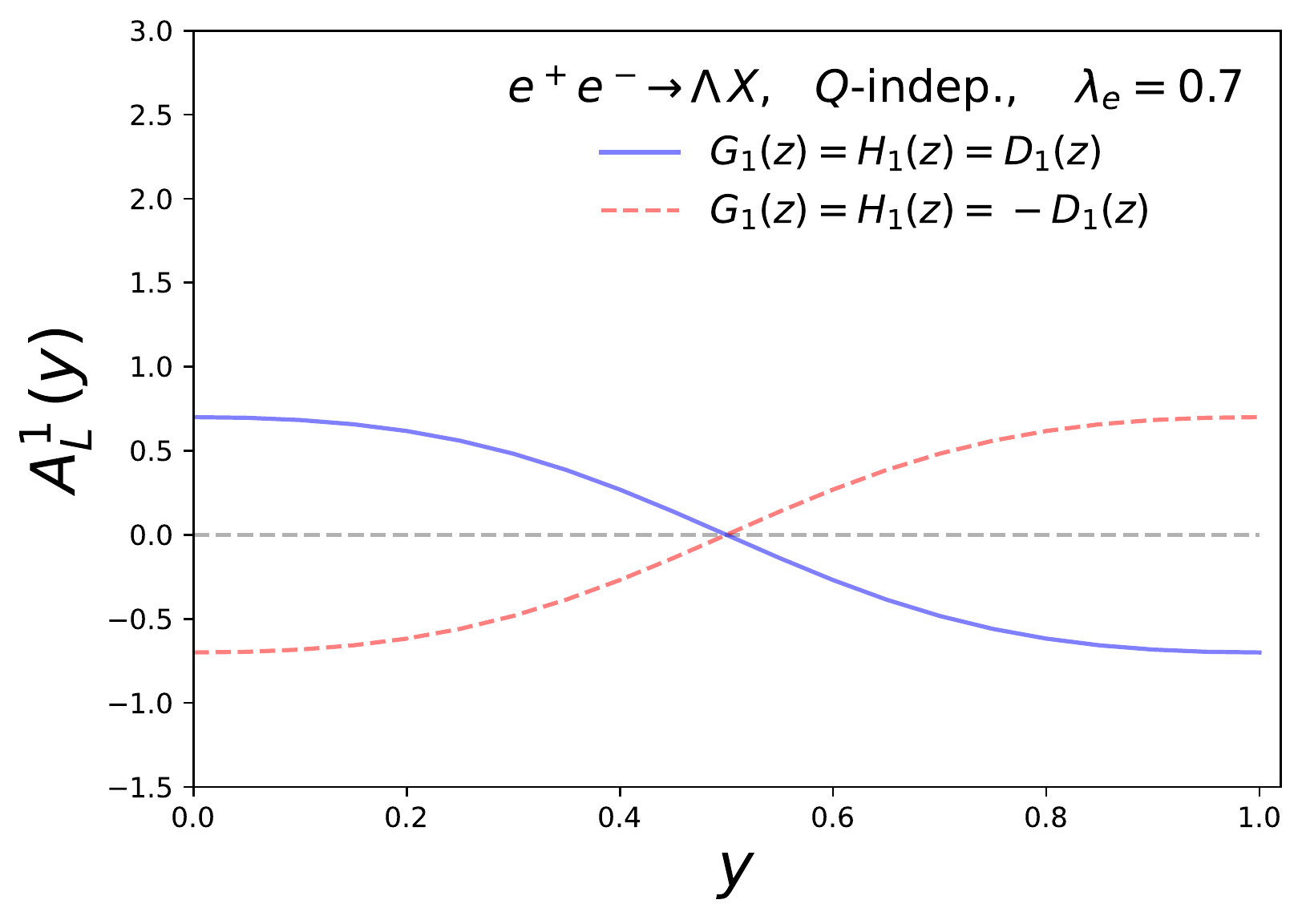}
	\includegraphics[width=0.49\linewidth]{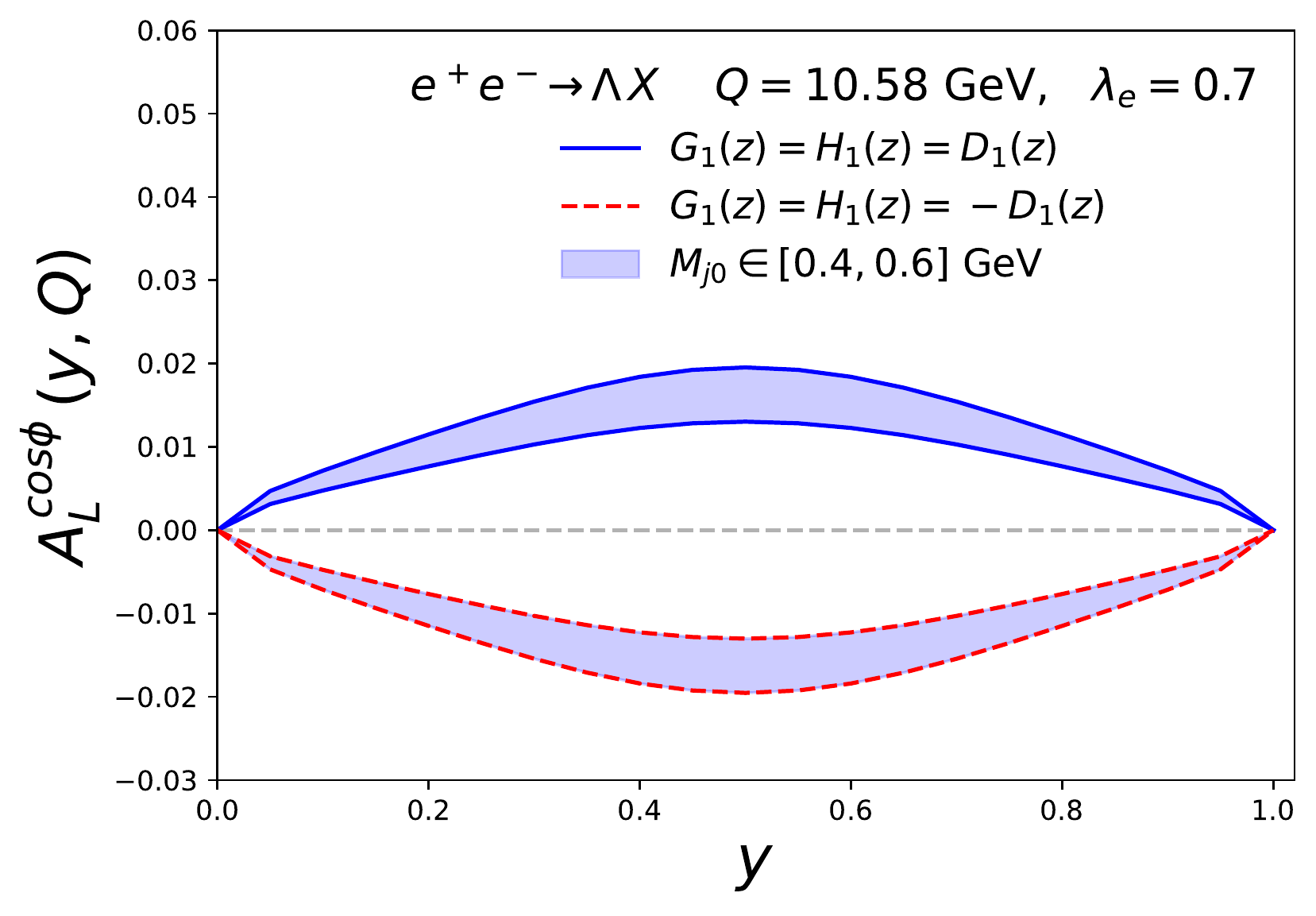}
	\caption{The Fourier components $A_L^1(y)$ and $A_L^{\cos\phi}(y,Q)$ of the longitudinal electron spin asymmetry as a function of $y$ at Belle-II's nominal energy $Q=10.58$ GeV. The band in the $\cos\phi$ modulation indicates the sensitivity of the measurement to $\pm20\%$ variation in the the jet mass at the initial scale.}
	\label{fig:AL_AT_Lambda}
\end{figure}

In Fig.~\ref{fig:AL_AT_Lambda} we then analyze two scenarios, in which the positivity bounds for $G_1$ and $H_1$ are saturated with opposite signs, $G_1(z) = H_1(z) = \pm D_1(z)$. 
Under the approximations discussed above and assuming that the jet mass is approximately flavor-independent, 
the contribution from the $D_1^\Lambda$ fragmentation function cancels in the asymmetry and we obtain
\begin{align}
\label{e:asymmetry_simpl}
    A_L(y,Q) 
    & = 
    \pm \underbrace{ \Big( \lambda_e\, \frac{C(y)}{2A(y)} \Big)}_{A_L^1(y)}  
    \lambda\, 
    \pm
    \underbrace{ \Big( 2\lambda_e\, \frac{M_j(Q)}{Q} \frac{D(y)}{A(y)}   \Big)}_{A_L^{\cos\phi}(y,Q)} |\bm{S}_T| \cos(\phi) \ ,
\end{align}
where $\lambda_e$ and $\lambda$ are the particle helicities, $\bm{S}_T$ is the transverse spin vector of the detected hadron.

The jet mass $M_j$ can then be extracted from the Fourier coefficient $A_L^{\cos\phi}$.
With the expected 70\% beam polarization at the polarized SuperKEKB upgrade~\cite{polarizedSuperKEKBWhitepaper}, the Fourier coefficient $A_L^{\cos\phi}$ is seen to be of ${\cal O}(1\%)$, reaching a maximum at $y=0.5$. 
At the same value of $y$ the constant modulation $A_L^1$ displays a node. This specific value allows one to separate the two modulations $A_L^1$ and $A_L^{\cos\phi}$, related to the longitudinal and transverse polarization of the detected hadron respectively. 
The blue band in Fig.~\ref{fig:AL_AT_Lambda} displays the sensitivity of this observable to a 20\% variation in the jet mass at the non-perturbative scale, $M_{j0} = 0.4-0.6$ GeV. 
The solid blue (dashed red) curves are related to the case where $G_1$ and $H_1$ saturate the positivity bound with a plus (minus) sign.

In summary, the $A_L^{\cos\phi}$ modulation of the beam spin asymmetry $A_L$ for production of a $\Lambda$ hadron in polarized $e^+e^-$ annihilation provides direct access to the dynamical component of the jet mass, allowing one to experimentally measure the contribution of the non-perturbative QCD dynamics at play in the hadronization mechanism. 
If the positivity bounds turn out not to be saturated, the signal may drop below the ${\cal O}(1\%)$ estimated above. 
However, being a twist-3 effect scaling as $\sim 1/Q$, the signal can significantly increase at lower center of mass energies.

Note that a similar jet mass measurement could be accomplished by detecting a di-hadron pair instead of a $\Lambda$ particle, with the pair's relative momentum playing the role of the $\Lambda$'s spin vector. A study of the same-side di-hadron production asymmetries is in progress.

\section{Acknowledgments}
This work was supported by the U.S. Department of Energy, Office of Science, Office of Nuclear Physics
under Award Numbers DE-SC0019230, DE-AC05-06OR23177, DE-SC0010007, DE-SC0008791, and the U.S. National Science Foundation under Nuclear Physics Award PHY-1913789.
F.~Ringer is supported by Simons Foundation 815892, NSF 1915093. 
P.A.~Gutierrez Garcia and I.~Scimemi are supported by the Spanish Ministry grant PID2019-106080GB-C21; and P.~Sanchez-Puertas by the Ministerio de Ciencia e Innovaci{\'o}n (grant PID2020-112965GB-I00). 
This project has received funding from the European Union Horizon 2020 research and innovation program under grant agreement Num. 824093 (STRONG-2020) and under grant agreement Num. 754510 (EU, H2020-MSCA-COFUND2016). 
S.~Leal Gomez is supported by the Austrian Science Fund FWF under the Doctoral Program W1252-N27 Particles
and Interactions.
A.~Signori acknowledges support from the European Commission through the Marie Sklodowska-Curie Action SQuHadron (grant agreement ID: 795475). 
T.~Sjöstrand is supported by the Swedish Research Council, contract number 2016-05996. 
This material is based upon work supported by the National Science Foundation Graduate Research Fellowship under Grant No.~NSF GRFP DGE 1644868.
\bibliography{snowMassBelleII}

\end{document}